\documentclass[a4paper,11pt]{article}
\usepackage{amsmath}
\usepackage{amsfonts}
\usepackage{latexsym}
\usepackage{graphicx}
\usepackage{multicol}
\usepackage{times}
\usepackage{latexsym}
\usepackage{enumerate}
\usepackage{epsfig}
\usepackage{amssymb}
\usepackage{amsfonts}
\usepackage[english]{babel}
\usepackage{graphics}
\usepackage{amsbsy}
\usepackage{setspace}
\usepackage{placeins}
\usepackage{float}
\usepackage{authblk}
\usepackage{multirow}
\usepackage{hyperref}
\usepackage{enumitem}
\usepackage{booktabs}
\usepackage[letterpaper, margin=1in]{geometry}
\newcommand\numberthis{\addtocounter{equation}{1}\tag{\theequation}} 
\begin{document}
\author[1]{Roberto Rocci\thanks{roberto.rocci@uniroma2.it}}
\author[2]{Stefano Antonio Gattone\thanks{gattone@unich.it}}
\author[1]{Roberto Di Mari\thanks{roberto.di.mari@uniroma2.it}}
\title{A data driven equivariant approach to constrained Gaussian mixture modeling}
\affil[1]{DEF, University of Rome Tor Vergata, Italy}
\affil[2]{DiSFPEQ, University G. d'Annunzio, Chieti-Pescara, Italy}
\date{}
\maketitle

\maketitle

\begin{abstract}
Maximum likelihood estimation of Gaussian mixture models with different class-specific covariance matrices is known to be problematic. This is due to the unboundedness of the likelihood, together with the presence of spurious maximizers. Existing methods to bypass this obstacle are based on the fact that unboundedness is avoided if the eigenvalues of the covariance matrices are bounded away from zero. This can be done imposing some constraints on the covariance matrices, i.e. by incorporating {\em a priori} information on the covariance structure of the mixture components. The present work introduces a constrained equivariant approach, where the class conditional covariance matrices are shrunk towards a pre-specified matrix $\boldsymbol{\Psi}.$ Data-driven choices of the matrix $\boldsymbol{\Psi},$ when {\em a priori} information is not available, and the optimal amount of shrinkage are investigated. The effectiveness of the proposal is evaluated on the basis of a simulation study and an empirical example.
{\bf keywords}: model based clustering, Gaussian mixture models, equivariant estimators.
\end{abstract}

\section{Introduction}
Let $\mathbf{x}$ be a $J$-variate random variable following a Gaussian mixture model (GMM) with density 
\begin{equation}
\label{md}
h(\mathbf{x};\theta)=\sum_{g=1}^G p_g \phi(\mathbf{x};\boldsymbol{\mu}_g,\boldsymbol{\Sigma}_g),
\end{equation}
where the $p_g$'s are the mixing proportions, with $p_g > 0$ $\forall g$ and $\sum_g p_g=1$, and
the component $\phi(\mathbf{x};\boldsymbol{\mu}_g,\boldsymbol{\Sigma}_g)$
represents the density of a $J$-variate normal distribution
with mean vector $\boldsymbol{\mu}_g$ and covariance matrix $\boldsymbol{\Sigma}_g$; furthermore, let us indicate the set of model parameters with $\theta=\left\{p_g, \boldsymbol{\mu}_g,\boldsymbol{\Sigma}_g \right\}_G=\left\{p_1,\dots p_G, \boldsymbol{\mu}_1,\dots,\boldsymbol{\mu}_G,\boldsymbol{\Sigma}_1,\dots,\boldsymbol{\Sigma}_G \right\} \in \mathbb{R}^{G(1+J+J^{2})},$
and the parameter space with  
\begin{equation}
\label{ps}
\Theta=\left\{\theta \in \mathbb{R}^{G(1+J+J^{2})} :\sum_g p_g=1, p_g > 0, \boldsymbol{\Sigma}_g \succ \boldsymbol{0},
g=1, \dots, G\right\},
\end{equation}
where the symbol $\succ$ refers to L\"{o}wner ordering on symmetric matrices and, in this case, is equivalent to requiring that $\boldsymbol{\Sigma}_g$ be positive definite.
The GMM is frequently used to classify a sample of observations. The idea is to consider the sample as drawn from a heterogeneous population where each sub-population is described by one component of the mixture. In other terms, each observation is assumed to come from one of the $G$ different groups characterized by the mixture components. 
The observations are classified into the groups by computing the posterior probabilities
\begin{equation}
p(g|\mathbf{x}) = \frac{p_g \phi(\mathbf{x};\boldsymbol{\mu}_g,\boldsymbol{\Sigma}_g)}
{\sum_h p_h \phi(\mathbf{x};\boldsymbol{\mu}_h,\boldsymbol{\Sigma}_h)},
\end{equation}
and assigning each observation to the group with the largest posterior probability.

The parameters of the GMM are generally unknown and estimated from the data. Given a sample of i.i.d. observations $\left\lbrace \mathbf{x}_i \right\rbrace_n= \left\{ \mathbf{x}_1, \mathbf{x}_2, \dots \mathbf{x}_n \right\}$, the estimation is usually done by maximizing the likelihood
\begin{equation}
\label{eq:like}
L\left(\theta; \left\lbrace \mathbf{x}_i \right\rbrace_n \right)=\prod_{i=1}^n  \left(\sum_{g=1}^G p_g \phi(\mathbf{x}_i;\boldsymbol{\mu}_g,\boldsymbol{\Sigma}_g)\right).
\end{equation}
The likelihood in Equation \eqref{eq:like} is known to be unbounded and it is cursed by the presence of several local maxima. As a consequence, the EM algorithm may fail to converge, leading to such degenerate solutions. To face degeneracy, several methods have been proposed by the literature in which constraints or penalties are added to the log-likelihood. Their main objective is to keep the eigenvalues of the class conditional covariance matrices bounded away from zero.

This paper considers the sufficient condition formulated by Ingrassia (2004) such that Hathaway's (1985) constraints hold: we propose a generalization that enforces the equivariance with respect to linear affine transformations of the data. The idea is to shrink the class conditional covariance matrices towards a pre-specified matrix $\boldsymbol{\Psi}.$ We investigate possible data-driven methods for choosing the matrix $\boldsymbol{\Psi},$ when {\em a priori} information on the group-specific covariance structure is not available, and we let the data determine the optimal amount of shrinkage. The equivariance property the method possesses is a key feature for twofold reasons. First, it means that irrespective of the kind of standardization performed on the data, the final clustering will be the same - provided that $\boldsymbol{\Psi}$ is transformed accordingly. Second, whatever the scale of the data is as they come in, there will be no \emph{best} pre-processing of the data ensuring a \emph{better} result, as the final clustering is not affected by changes in scale.   

The plan of the paper is the following. Section \ref{degen} gives insights on the notion of degeneracy for multivariate GMM, and Section \ref{remdegen} reviews some of the workarounds proposed by the existing literature. In Section \ref{invar} we state the property of equivariance of GMM and we show, in Section \ref{constr1}, that the property holds in the constrained approach of Hathaway (1985), whereas it does not hold in the sufficient condition provided by Ingrassia (2004). In Section \ref{constr2} we illustrate how these constraints can be generalized, by introducing the matrix $\boldsymbol{\Psi},$ to become equivariant under linear affine transformations of the data, provided that $\boldsymbol{\Psi}$ is transformed accordingly, and how their configuration can be tuned from the data (Section \ref{crossvalid}). Section \ref{alg} summarizes the algorithm. The proposal is evaluated through a simulation study (Section \ref{simulation}) and an empirical application (Section \ref{wineapp}). Section \ref{concl} concludes with a final discussion.
\section{The issue of degeneracy in GMM}\label{degen}
In the univariate case, the likelihood function increases without bound if some variances tend to zero and the corresponding component's mean coincides with a sample observation (Kiefer and Wolfowitz, 1956; Day, 1969). Biernacki and Chr\'etien (2003) showed that if mixture parameters are close to a degenerate solution, then the EM is attracted by it and the divergence is extremely fast. Although Kiefer (1978) proved that maximum likelihood does not fail, as there exists a local maximizer strongly consistent and asymptotically efficient, several local maximizers can exist for a given sample. That is, some local maximizers are spurious, i.e. with a high likelihood but of little practical use because highly biased. They are characterized by some component variances and mixing proportions very small relative to the others (Day, 1969; McLachlan and Peel, 2000). Detecting the desired solution, among the many available, can therefore be a complicated task.

The same problems hold in the multivariate case (as an example, see Ingrassia and Rocci, 2011, for an extension of Biernacki and Chr\'etien, 2003), with additional complications. To notice how unboundedness is caused, first of all let us express the density of the $i$-th observation on the $g$-th component as follows
\begin{equation}\label{Lispec}
\phi\left(\mathbf{x}_i;\boldsymbol{\mu}_g, \boldsymbol{\Sigma}_g \right) = \frac{1}{\sqrt{2\pi \prod_{j=1}^{J}\lambda_{jg}}} \exp\left\{ -\frac{1}{2} (\mathbf{x}_{i} - \boldsymbol{\mu}_{g})'\mathbf{Q}_{g}\mathbf{L}_{g}^{-1}\mathbf{Q}_{g}'(\mathbf{x}_{i} - \boldsymbol{\mu}_{g}) \right\},
\end{equation}
where $\mathbf{Q}_{g}$ is the square $J \times J$ matrix whose $j-$th column is the eigenvector $\mathbf{q}_{jg}$ of $\mathbf{\Sigma}_{g},$ and $\mathbf{L}_{g}$ is the diagonal matrix whose diagonal elements are the corresponding eigenvalues $\{\lambda_{jg}\}_{J},$ ordered such that $\lambda_{1g} \geq \dots \geq \lambda_{Jg}.$ Equation \eqref{Lispec} can be rewritten as
\begin{align*}
\phi\left(\mathbf{x}_i;\boldsymbol{\mu}_g, \boldsymbol{\Sigma}_g \right) &= \frac{1}{\sqrt{2\pi \prod_{j=1}^{J}\lambda_{jg}}} \exp\left\{ -\frac{1}{2} (\mathbf{x}_{i} - \boldsymbol{\mu}_{g})'(\sum_{j=1}^{J} \lambda_{jg}^{-1} \mathbf{q}_{jg} \mathbf{q}_{jg}')(\mathbf{x}_{i} - \boldsymbol{\mu}_{g}) \right\} \\
&= \frac{1}{\sqrt{2\pi \prod_{j=1}^{J}\lambda_{jg}}} \exp\left\{ -\frac{1}{2} \sum_{j=1}^{J}\lambda_{jg}^{-1}[(\mathbf{x}_{i} - \boldsymbol{\mu}_{g})'\mathbf{q}_{jg}][\mathbf{q}_{jg}'(\mathbf{x}_{i} - \boldsymbol{\mu}_{g})] \right\} \\
&= \frac{1}{\sqrt{2\pi \prod_{j=1}^{J}\lambda_{jg}}} \exp\left\{ -\frac{1}{2} \sum_{j=1}^{J}\lambda_{jg}^{-1}[(\mathbf{x}_{i} - \boldsymbol{\mu}_{g})'\mathbf{q}_{jg}]^{2} \right\}. \numberthis \label{eq:Li2parts}
\end{align*}
As Policiello (1981) argued, the likelihood in Equation \eqref{eq:like} can be written as the sum of non negative terms. 
Among them, it is possible to isolate the product of the density of the $i$-th observation on 
the $g$-th component - Equation \eqref{eq:Li2parts} - and the densities of the other observations on the other components and the corresponding mixing proportions. If observation $i$ is such that $\mathbf{x}_{i}'\mathbf{q}_{Jg} - \boldsymbol{\mu}_{g}'\mathbf{q}_{Jg} = 0,$ then, as $\lambda_{Jg} \rightarrow 0,$ there would be no exponential term involving $\lambda_{Jg}$ who can attenuate the effect of $\frac{1}{\sqrt{2\pi \prod_{j=1}^{J}\lambda_{jg}}} \rightarrow \infty.$ In words, the sample likelihood diverges when in one component the covariance matrix is close to singularity and the projection of the component's mean on the eigenvector corresponding to the smallest eigenvalue coincides with the projection of one of the observations on the same eigenvector. 

\section{Remedies to degeneracy}\label{remdegen}  
The easiest way to handle degeneracy is to initialize the EM algorithm from several starting points until a local maximum is found (Biernacki and Chr\'etien, 2003). McLachlan and Peel (2000) proposed monitoring the local maximizers by inspecting the relative size of the estimated mixing proportions and component variances. This leads, in practice, to performing maximum likelihood estimation by looking for the correct local maximum and discarding those that seem to be spurious.

Further methods exploit constraints on the covariance matrices. This approach is based on the seminal work of Hathaway (1985), where he studied how to avoid the divergence of the likelihood in the univariate case by imposing a lower bound, say $c$, to the ratios of the scale parameters. In this way the variances cannot be arbitrarily different. Hathaway proved the boundedness of the likelihood  and the consistency of the ML estimator under such constraints. In the multivariate case, the lower bound is imposed on the generalized eigenvalues of each pair of covariance matrices and the ML estimator results to be equivariant under linear affine transformations of the data. This implies that, as in the unconstrained case, if the data are linearly transformed, the estimated posterior probabilities do not change and the clustering remains unaltered (see Sections \ref{invar} and \ref{constr1}).

An important issue is the choice of the constant $c,$ which controls the strength of the constraints. In the context of univariate mixtures of Gaussians or linear regression models, some authors have shown that the maximum likelihood constrained estimator is consistent if $c$ decreases to zero at a certain rate as the sample size increases to infinity (e.g. Tanaka and Takemura (2006), Tan et al. (2007), Xu et al. (2010)). Nevertheless, finite-sample sensible choice of $c$ is still an open issue.

Hathaway's constraints are very difficult to apply within iterative procedures like the EM algorithm. To solve this problem, Ingrassia (2004) proposed to simplify the constraints by putting bounds on the eigenvalues of the covariance matrices. Although putting lower bounds on the group conditional covariance matrices was already common practice, Ingrassia (2004) found a way to reconcile Hathaway's contribution with the common practice: his bounds on the eigenvalues give a sufficient condition such that Hathaway's constraints are satisfied. The simplification is such that the constraints can be easily implemented within the EM algorithm, preserving its monotonicity property (as shown in Ingrassia and Rocci, 2007).

Several authors extended the constrained setup of Ingrassia (2004). Greselin and Ingrassia (2013) applied this setup to mixtures of factor analyzers. They proposed a tuning procedure for selecting the bounds for the eigenvalues of the covariance matrices, based on the final likelihood over a set of runs. Ingrassia and Rocci (2011) modified the constrained algorithm, allowing for stringent constraints which are lifted during the iterations. Browne et al. (2013) combined the ideas in Ingrassia and Rocci (2007, 2011), constraining dynamically the smallest eigenvalue, the largest eigenvalue and both the smallest and the largest ones. All of these proposals share the drawback of not being affine equivariant.

Gallegos and Ritter (2009a; 2009b), and Ritter (2014) applied Hathaway's constraints to robust clustering. They proposed to obtain all local maxima of the trimmed likelihood and, for each solution, investigate the value of $c$ such that it fulfills the constraints. The idea is to choose, \emph{a posteriori}, the solution  with the highest trade-off between scale balance ($c$) and fit (log-likelihood). This approach can be viewed as a refined version of what was proposed in McLachlan and Peel (2000). Garcia-Escudero et al. (2008), from the same strand of literature, introduced the TCLUST algorithm, based on controlling the relative sizes  of the eigenvalues of the cluster scatter matrices. The TCLUST algorithm implies solving several complex optimization problems. Fritz et al. (2013) and Garcia-Escudero et al. (2014) proposed further improvements to the algorithm in order to make it more efficient. The constraints considered therein are not affine equivariant.

Seo and Kim (2012) pointed out that singular and spurious solutions overfit random localized patterns composed of few observations in the dataset. Such observations have a strong influence on the formation of the likelihood-based solutions. Their proposal was to take out such, say, $k$ observations with the highest likelihood (likelihood-based $k$-deleted method), or with the highest value for a score-based statistic (score-based $k$-deleted method). In this way the likelihood of the reduced samples is evaluated at each local maximizer previously found: the root they suggested to select is the one with the highest $k$-deleted likelihood. Kim and Seo (2014) show that their score-based method can be fairly well approximated with the computationally more efficient gradient-based version of the $k$-deleted method. 

The degeneracy problem may also be addressed by  adding a penalty to the log-likelihood (penalized approach). Ciuperca et al. (2003) have shown the consistency of the penalized likelihood estimators proposed in Ridolfi and Idier (1999, 2000) for univariate GMM. Chen and Tan (2009) extended the consistency result for the multivariate case. In this framework, the penalty term on the component covariance is added to the log-likelihood (Snoussi and Djafari, 2001; Chen et al, 2008). This penalty can be interpreted as the log of the prior distribution in a Maximum-A-Posteriori estimation setup. Yet, the penalized methods are not affine equivariant, unless the prior's hyperparameters are suitably transformed. MAP estimation, with an \emph{a priori} distribution for the covariance matrices, is what Fraley and Raftery (2007) suggested to use, instead of Maximum-Likelihood, to circumvent the issues of degeneracy and spurious solutions.

\section{Equivariance in the Gaussian Mixture model}\label{invar}

The maximum likelihood estimators (MLE) of Equation (\ref{md}) are equivariant with respect to linear affine transformations of the data. That is, if the data are linearly transformed, the MLE are transformed accordingly. This property is particularly important
in classification because it implies that linear affine transformations of the data 
do not change the posterior estimates (Kleinberg, 2002; Ritter, 2014).
 
The equivariance property can be shown in the following way.
Let us define a linear affine transformation $\mathbf{x}^*=\mathbf{Ax}+\mathbf{b}$, where $\mathbf{A}$ is non singular.
It is well known that
\begin{align*}
\phi(\mathbf{x};\boldsymbol{\mu},\boldsymbol{\Sigma}) & = |\mathbf{A}|\phi(\mathbf{Ax}+\mathbf{b};
\mathbf{A}\boldsymbol{\mu}+\mathbf{b},\mathbf{A}\boldsymbol{\Sigma}\mathbf{A}')\\
& =  |\mathbf{A}|\phi(\mathbf{x}^*;\boldsymbol{\mu}^*,
\boldsymbol{\Sigma}^*) \numberthis  \label{eq:equiGauss}
\end{align*}
where $\boldsymbol{\mu}^*=\mathbf{A}\boldsymbol{\mu}+\mathbf{b}$ and $\boldsymbol{\Sigma}^*=\mathbf{A}\boldsymbol{\Sigma}\mathbf{A}'$.
 This implies that, denoting the likelihood of the original data with $\mathcal{F}$ and the likelihood of the transformed data with $\mathcal{F}^*$, we have, with obvious notation
\begin{align*}
\mathcal{F} & =  L\left(\left\lbrace p_g,\boldsymbol{\mu}_g, \boldsymbol{\Sigma}_g \right\rbrace_G; \left\lbrace \mathbf{x}_i \right\rbrace_n  \right) = \prod_{i=1}^n \sum_{g=1}^G  p_g \phi(\mathbf{x}_i;\boldsymbol{\mu}_g,\boldsymbol{\Sigma}_g) \\
& =   \prod_{i=1}^n \sum_{g=1}^G  p_g |\mathbf{A}| \phi(\mathbf{x}_i^*;\boldsymbol{\mu}_g^*,\boldsymbol{\Sigma}_g^*) =
|\mathbf{A}|^n L\left(\left\lbrace p_g,\boldsymbol{\mu}_g^*, \boldsymbol{\Sigma}_g^* \right\rbrace_G; \left\lbrace \mathbf{x}_i^* \right\rbrace_n \right)\\
& =  |\mathbf{A}|^n \mathcal{F}^*. \numberthis \label{eq:equiGaussmix}
\end{align*}
It follows that there exists a one to one correspondence among the local maxima of $\mathcal{F}$ and $\mathcal{F}^*$. 
In particular, if 
$\left\{\hat{p}_g,\hat{\boldsymbol{\mu}}_g, \hat{\boldsymbol{\Sigma}}_g \right\}_G$ is a local maximizer for $\mathcal{F},$ then $\left\{\hat{p}_g,\mathbf{A}\hat{\boldsymbol{\mu}}_g+\mathbf{b}, \mathbf{A}\hat{\boldsymbol{\Sigma}}_g \mathbf{A}' \right\}_G$ will be a local maximizer for $\mathcal{F}^*$. Analogously, if $\left\{\hat{p}_g^*,\hat{\boldsymbol{\mu}}_g^*, \hat{\boldsymbol{\Sigma}}_g^* \right\}_G$ is a local maximizer for $\mathcal{F}^*$, then 
$\left\{\hat{p}_g^*,\mathbf{A}^{-1}(\hat{\boldsymbol{\mu}}_g^*-\mathbf{b}), \mathbf{A}^{-1}\hat{\boldsymbol{\Sigma}}_g^* (\mathbf{A}')^{-1} \right\}_G$ 
will be a local maximizer for $\mathcal{F}$.
It is interesting to note that every pair of local maximizers produces the same estimates of the posterior probabilities, that is
\begin{eqnarray*}
\hat{p}(g|\mathbf{x}_i) & = &\frac{\hat{p}_g \phi(\mathbf{x}_i;\hat{\boldsymbol{\mu}}_g,\hat{\boldsymbol{\Sigma}}_g)}{\sum_h \hat{p}_h \phi(\mathbf{x}_i;\hat{\boldsymbol{\mu}}_h,\hat{\boldsymbol{\Sigma}}_h)}\\
& = & \frac{\hat{p}_g |\mathbf{A}| \phi(\mathbf{x}_i^*;\hat{\boldsymbol{\mu}}_g^*,\hat{\boldsymbol{\Sigma}}_g^*)}{\sum_h \hat{p}_h |\mathbf{A}| \phi(\mathbf{x}_i^*;\hat{\boldsymbol{\mu}}_h^*,\hat{\boldsymbol{\Sigma}}_h^*)}=
\hat{p}^*(g|\mathbf{x}^*_i).
\end{eqnarray*}

The above equality proves that the classification obtained via the GMM model is invariant under the group of linear affine transformations on the data $\left\lbrace \mathbf{x}_i\right\rbrace_n$. 
This property is crucial when dealing with practical applications as it implies that the clustering does not depend on the choice of a particular method of data standardization - which could instead affect the inference.

\section{Constraints on covariance eigenvalues}\label{constr1}
Hathaway (1985) proposed to impose the following restrictions on the covariance matrices
\begin{equation}
\label{ch}
\lambda_j (\boldsymbol{\Sigma}_g\boldsymbol{\Sigma}_h^{-1}) \geq c,
\hspace{1cm}j=1,\dots,J;\hspace{0.2cm} g,h=1 \dots G
\end{equation}
where $\lambda_j(\mathbf{A})$ is the $j$-th eigenvalue of $\mathbf{A}$ and $0<c\leq 1$.
This prevents the likelihood from diverging and reduces the number of spurious maximizers. However, the method is difficult to implement and a correct choice of $c$ is not simple in practice.  A value of $c$ close to $1$ could exclude the correct solution, whereas a value too close to $0$ is likely to increase the chance of converging to a spurious maximizer.

Ingrassia (2004) simplified Hathaway's constraints as
\begin{equation} 
\label{nc}
\sqrt{c} \leq \lambda_j (\boldsymbol{\Sigma}_g)\leq \frac{1}{\sqrt{c}}, \hspace{1cm}j=1,\dots,J;\hspace{0.2cm} g=1 \dots G.
\end{equation}
It is easy to show that (\ref{nc}) implies Hathaway's constraints (\ref{ch}) while the reverse is not necessarily true (Ingrassia, 2004). This ensures a bounded likelihood, and a reduction in the number of spurious maximizers. The constraints are easy to implement, as shown in  Rocci and Ingrassia (2007); however, choosing an optimal \textit{c} is still an issue. 

It is important to check if the above constrained approaches offer equivariant estimators under linear affine transformations.
The property can be shown to hold for Hathaway's approach as follows.

Let $\left\lbrace \mathbf{x}_i \right\rbrace_n= \left\{ \mathbf{x}_1, \mathbf{x}_2, \dots \mathbf{x}_n \right\}$ be a sample of i.i.d. observations. The estimates are computed as the solution of the optimization problem
\begin{equation}\label{eq:maxcon1}
\begin{cases}
 \max\limits_{\left\lbrace p_g,\boldsymbol{\mu}_g, \boldsymbol{\Sigma}_g \right\rbrace_G}& L\left(\left\lbrace p_g,\boldsymbol{\mu}_g, \boldsymbol{\Sigma}_g \right\rbrace_G; \left\lbrace \mathbf{x}_i \right\rbrace_n  \right) \\
 \qquad \text{s.t.} & \lambda_j (\boldsymbol{\Sigma}_g\boldsymbol{\Sigma}_h^{-1}) \geq c,\hspace{1cm}j=1,\dots,J;\hspace{0.2cm} g,h=1 \dots G. 
\end{cases}
\end{equation}
Given the transformation $\mathbf{x}^*=\mathbf{Ax}+\mathbf{b},$ the maximand in Equation (\ref{eq:maxcon1}) can be rewritten (see Section \ref{invar}) as 
\[
L\left(\left\lbrace p_g,\boldsymbol{\mu}_g, \boldsymbol{\Sigma}_g \right\rbrace_G; \left\lbrace \mathbf{x}_i \right\rbrace_n  \right) 
=|\mathbf{A}|^n L\left(\left\lbrace p_g,\boldsymbol{\mu}_g^*, \boldsymbol{\Sigma}_g^* \right\rbrace_G; \left\lbrace \mathbf{x}_i^* \right\rbrace_n \right),
\]
where $\boldsymbol{\mu}^*=\mathbf{A}\boldsymbol{\mu}+\mathbf{b}$ and $\boldsymbol{\Sigma}^*=\mathbf{A}\boldsymbol{\Sigma}\mathbf{A}'$.

Noting that  
\begin{eqnarray*}
\lambda_j (\boldsymbol{\Sigma}_g \boldsymbol{\Sigma}_h^{-1})
&=&
\lambda_j(\mathbf{A}^{-1}\mathbf{A}\boldsymbol{\Sigma}_g \mathbf{A}'(\mathbf{A}')^{-1} \boldsymbol{\Sigma}_h^{-1})\\
&=&
\lambda_j(\mathbf{A}\boldsymbol{\Sigma}_g \mathbf{A}'(\mathbf{A}')^{-1} \boldsymbol{\Sigma}_h^{-1}\mathbf{A}^{-1})\\
&=&
\lambda_j (\boldsymbol{\Sigma}_g^* (\boldsymbol{\Sigma}_h^*)^{-1}),
\end{eqnarray*}
we can equivalently write the optimization problem in Equation (\ref{eq:maxcon1}) as
\begin{equation}\label{eq:maxcon12}
\begin{cases}
 \max\limits_{\left\lbrace p_g,\boldsymbol{\mu}_g^*, \boldsymbol{\Sigma}_g^* \right\rbrace_G}& L\left(\left\lbrace p_g,\boldsymbol{\mu}_g^*, \boldsymbol{\Sigma}_g^* \right\rbrace_G; \left\lbrace \mathbf{x}_i^* \right\rbrace_n \right) \\
 \qquad \text{s.t.} & \lambda_j (\boldsymbol{\Sigma}_g^* (\boldsymbol{\Sigma}_h^*)^{-1}) \geq c,\hspace{1cm}j=1,\dots,J;\hspace{0.2cm} g,h=1 \dots G. 
\end{cases}
\end{equation}
It follows that if $\left\{\hat{p}_g,\hat{\boldsymbol{\mu}}_g, \hat{\boldsymbol{\Sigma}}_g \right\}_G$ is a maximizer for (\ref{eq:maxcon1}), then $\left\{\hat{p}_g,\mathbf{A}\hat{\boldsymbol{\mu}}_g+\mathbf{b}, \mathbf{A}\hat{\boldsymbol{\Sigma}}_g \mathbf{A}' \right\}_G = \left\{\hat{p}_g,\hat{\boldsymbol{\mu}}^*_g, \hat{\boldsymbol{\Sigma}}^*_g \right\}_G$ is a maximizer for (\ref{eq:maxcon12}) and vice-versa, and the two maximization problems are equivalent. As in the unconstrained case, every pair of local maximizers produces the same estimates of the posterior probabilities. 
This property does not hold for the constraints given in (\ref{nc}). That is, if $\left\lbrace \hat{\boldsymbol{\Sigma}}_g \right\rbrace_G$ is a constrained local maximizer for $\mathcal{F}$ subject to (\ref{nc}), $\left\lbrace\mathbf{A}\hat{\boldsymbol{\Sigma}}_g \mathbf{A}'\right\rbrace_{G}$ does not necessarily satisfy (\ref{nc}). As an example, let us suppose that $s_{max}(\mathbf{A})< \sqrt{c},$ where $s_{max}(\mathbf{A})$ is the largest singular value of $\mathbf{A}$. In this case, for a given $g,$
\begin{eqnarray*}
\lambda_j(\mathbf{A}\hat{\boldsymbol{\Sigma}}_g \mathbf{A}') &\leq &  \lambda_{max}(\mathbf{A}\hat{\boldsymbol{\Sigma}}_g \mathbf{A}')
\leq s_{max}(\mathbf{A})^2 \lambda_{max}(\hat{\boldsymbol{\Sigma}}_g)\\ 
& \leq &s_{max}(\mathbf{A})^2 \frac{1}{\sqrt{c}} < c \frac{1}{\sqrt{c}}=\sqrt{c}.
\end{eqnarray*}
We conclude that $\mathbf{A}\hat{\boldsymbol{\Sigma}}_g \mathbf{A}'$ does not satisfy the constraints in (\ref{nc}) because $\lambda_j(\mathbf{A}\hat{\boldsymbol{\Sigma}}_g \mathbf{A}') < \sqrt{c}$, and then it cannot be a constrained local maximizer for $\mathcal{F}^*$.

Constrains in (\ref{nc}) are such that there is no one to one correspondence between the set of local maximizers of $\mathcal{F}$ and $\mathcal{F}^*$.   
Thus, the method suffers the disadvantage that the clustering depends on the choice of matrix $\mathbf{A}$. To fix this, data standardization is not the best way to go for two main reasons. First, the standardization requires a choice for the matrix $\mathbf{A}$ and, second, there is no single best approach to data standardization (Milligan and Cooper, 1988; Doherty \textit{et al}, 2007).

It is now clear that affine equivariance is not just a desirable property. It is one of the basic requirements of any clustering method, which should not be sensitive to the changes in the units of measurement of the data. In the next section, our goal will be that of deriving a new set of constraints that are affine equivariant. With an affine equivariant clustering method, researchers and practitioners shall not be concerned anymore with choosing what method to adopt to standardize their data.
 \section{Equivariant constraints} \label{constr2}
 Our proposal is to generalize the constraints (\ref{nc}) by
\begin{equation}
\label{eq:gnc}
\sqrt{c} \leq \lambda_j (\boldsymbol{\Sigma}_g\boldsymbol{\Psi}^{-1})\leq \frac{1}{\sqrt{c}},
\hspace{1cm}j=1,\dots,J;\hspace{0.2cm} g=1 \dots G
\end{equation}
where $\boldsymbol{\Psi}$ is a symmetric positive definite matrix representing our \textit{prior} information about the covariance structure. Clearly, \eqref{eq:gnc} is equal to (\ref{nc}) when $\boldsymbol{\Psi}=\mathbf{I}$.

It can be shown that the above constraints imply Hathaway's constraints. 
It is known that (Anderson and Gupta, 1963)
\begin{equation}
\lambda_{min}(\mathbf{AB}^{-1})\geq \lambda_{min}(\mathbf{AC}^{-1})
\lambda_{min}(\mathbf{CB}^{-1}),
\end{equation}
where $\mathbf{A}$ is a positive semi-definite matrix and $\mathbf{B}$ and $\mathbf{C}$ are positive definite matrices.
Now, if \eqref{eq:gnc} holds, then
\begin{equation}
\label{re}
\lambda_{min}(\boldsymbol{\Sigma}_g \boldsymbol{\Sigma}_h^{-1}) \geq 
\lambda_{min}(\boldsymbol{\Sigma}_g\boldsymbol{\Psi}^{-1})
\lambda_{min}(\boldsymbol{\Psi} \boldsymbol{\Sigma}_h^{-1})=
\frac{\lambda_{min}(\boldsymbol{\Sigma}_g\boldsymbol{\Psi}^{-1})}
{\lambda_{max}(\boldsymbol{\Sigma}_h \boldsymbol{\Psi}^{-1})}\geq \frac{\sqrt{c}}{\frac{1}{\sqrt{c}}}= c.
\end{equation}
Thus, \eqref{eq:gnc} implies (\ref{ch}). 
 
Furthermore, it can be shown that \eqref{eq:gnc} is invariant under linear and affine transformations provided that $\boldsymbol{\Psi}$ is transformed accordingly, i.e. it is replaced by $\boldsymbol{\Psi}^*=\mathbf{A}\boldsymbol{\Psi} \mathbf{A}'$.
If $\left\lbrace \hat{\boldsymbol{\Sigma}}_g \right\rbrace_G$ is a constrained local maximizer for $\mathcal{F}$ subject to \eqref{eq:gnc}, then $\left\lbrace \hat{\boldsymbol{\Sigma}}^*_g = \mathbf{A}\hat{\boldsymbol{\Sigma}}_g \mathbf{A}' \right\rbrace_G$ is a local maximizer for $\mathcal{F}^*$ subject to \eqref{eq:gnc} for $g=1 \dots G$. We have that
\begin{eqnarray*}
\label{eiglt}
\lambda_j (\hat{\boldsymbol{\Sigma}}_g \boldsymbol{\Psi}^{-1}) &=&
\lambda_j  (\hat{\boldsymbol{\Sigma}}_g\mathbf{A}'(\mathbf{A}')^{-1} \boldsymbol{\Psi}^{-1}\mathbf{A}^{-1}\mathbf{A})\\
&=&\lambda_j(\mathbf{A}\hat{\boldsymbol{\Sigma}}_g \mathbf{A}'(\mathbf{A}')^{-1} \boldsymbol{\Psi}^{-1}\mathbf{A}^{-1}) \\
&=& \lambda_j (\hat{\boldsymbol{\Sigma}}^*_g \boldsymbol{(\Psi^*)}^{-1}).
\end{eqnarray*}

In words, if a linear affine transformation is performed on the data, $\boldsymbol{\Psi}$ must be changed accordingly. This scheme of transforming $\boldsymbol{\Psi}$ ensures the equivariance of the method. By contrast, holding $\boldsymbol{\Psi}$ fixed breaks the equivariance property.

The constraints \eqref{eq:gnc} have the effect of shrinking the covariance matrices to $\boldsymbol{\Psi},$ and the level of shrinkage is given by the value of $c.$ Note that for
$c=1$, $\hat{\boldsymbol{\Sigma}}_g = \boldsymbol{\Psi},$ whereas for $c\rightarrow 0,$ $\hat{\boldsymbol{\Sigma}}_g$ equals the unconstrained ML estimate. Furthermore we can show that the Stein's discrepancy - known as Stein's loss (James and Stein, 1961) - between the matrices $\hat{\boldsymbol{\Sigma}}_g$ and $\boldsymbol{\Psi}$ goes to zero as $c$ approaches one. The Stein's discrepancy between the matrices $\hat{\boldsymbol{\Sigma}}_g$ and $\boldsymbol{\Psi}$ is
\begin{equation}\label{eq:Lstein1}
\text{L}(\hat{\boldsymbol{\Sigma}}_g,\boldsymbol{\Psi}) = \text{tr}(\hat{\boldsymbol{\Sigma}}_g\boldsymbol{\Psi}^{-1}) - \log |\hat{\boldsymbol{\Sigma}}_g\boldsymbol{\Psi}^{-1}| - J \geq 0.
\end{equation}
Let us rewrite Equation \eqref{eq:Lstein1} as follows.
\begin{equation}\label{eq:Lstein2}
\text{L}(\hat{\boldsymbol{\Sigma}}_g,\boldsymbol{\Psi}) = \sum_{j=1}^{J} \lambda_{j} (\hat{\boldsymbol{\Sigma}}_g\boldsymbol{\Psi}^{-1}) - \sum_{j=1}^{J} \log(\lambda_{j} (\hat{\boldsymbol{\Sigma}}_g\boldsymbol{\Psi}^{-1})) - J  
\end{equation}
Using the constraints in \eqref{eq:gnc}, we can derive the following majorizing function
\begin{equation}\label{eq:minorz}
\text{L}(\hat{\boldsymbol{\Sigma}}_g,\boldsymbol{\Psi}) \leq \frac{J}{\sqrt{c}} - J\log(\sqrt{c}) - J,
\end{equation}
which is decreasing in $c.$ This can be shown by noting that the first derivative of the right-hand side of \eqref{eq:minorz} with respect to $c$ is equal to $-\frac{J}{2c\sqrt{c}} - \frac{J}{2c},$ and is negative when $0<c\leq 1$. This implies that the function is decreasing when $c$ increases within the interval $(0,1]$.

Intuitively, the constraints \eqref{eq:gnc} provide with a way to obtain a model in between a too restrictive model, the homoscedastic, and an ill-conditioned model, the heteroscedastic. 

\section{Data-driven choice of $\boldsymbol{\Psi}$ and $c$} \label{crossvalid}

Issues arise when \emph{a priori} information about the structure of the class conditional covariance matrices is not available. In that case, $\boldsymbol{\Psi}$ and $c$ have to be selected from the data. From the previous discussion, for a given $c,$ every $\hat{\boldsymbol{\Sigma}}_g$ cannot be too far from $\boldsymbol{\Psi}$ in terms of Stein's discrepancy. Thus $\boldsymbol{\Psi}$ can be seen as the barycenter of the cloud of the $\hat{\boldsymbol{\Sigma}}_g$'s: the \emph{average} conditional covariance matrix. Therefore, the most natural choice is to estimate such \emph{average} as the within covariance matrix of the homoscedastic Gaussian model. How close the final clustering will be to the homoscedastic model will depend on the value of the tuning constant: for values of $c$ close to $0$, the resulting clustering will be close to that of the heteroscedastic mixture model, whereas $c\rightarrow 1$ implies a clustering close to that of the homoscedastic mixture model. 

Other possible choices of $\boldsymbol{\Psi},$ which guarantee the equivariance of the constraints, are available: the sample covariance matrix, which is computationally faster and is frequently used as hyperparameter in Bayesian Gaussian mixtures (for instance, see Fraley and Raftery, 2007), or the within covariance matrix of a homoscedastic mixture of Student-\textit{t}. To motivate this, let us recall that a random vector conditionally distributed as a multivariate Gaussian, given Wishart inverse covariance matrix, has a multivariate Student-$t$ distribution (Dawid, 1981; Dickey, 1967). Using similar arguments as in Peel and McLachlan (2000), if $\mathbf{x}| \boldsymbol{\Sigma}_1, \dots, \boldsymbol{\Sigma}_G$ is a GMM, and $\boldsymbol{\Sigma}_1^{-1}, \dots, \boldsymbol{\Sigma}_G^{-1}$ are i.i.d. Wishart random variables, the marginal distribution of $\mathbf{x}$ is a homoscedastic mixture of Student-$t$'s.

The choice of $c$ is crucial. A value of $c$ too large could exclude the right solution, whereas a too small value of $c$ is likely to increase the chance to converge to spurious local maxima: such solutions overfit random localized pattern composed of few data points being almost co-planar (Ritter, 2014; Seo and Kim, 2012). Hence, selecting $c$ jointly with the mixture parameters by maximizing the likelihood on the entire sample would trivially yield a scale balance approaching zero. 

A practical alternative would be to split the data into a training set, where model parameters are estimated, and a test set, where the log-likelihood is evaluated for a given value of $c.$ The optimal tuning parameter $c$ would then be selected such that the test set log-likelihood is maximized.

The use of the test set log-likelihood as a model selection tool is advocated by Smyth (1996; 2000), in the context of estimating the number of mixture components. The motivation behind its use is that it can be showed to be an unbiased estimator (within a constant) of the Kullback-Leibler divergence between the \emph{truth} and the model under consideration (Smyth, 2000). This means that, even under a misspecified model, the procedure renders a $c$ such that the Kullback-Leibler divergence is minimized.  

In spite of the usual unavailability of large independent test sets, a valid alternative is to use the \emph{cross-validated} log-likelihood in order to estimate the test set log-likelihood. This consists in repeatedly partitioning the data into training and test sets and, for a given $c,$ estimate the mixture parameters on the training sets. The model fit is then measured summing the log-likelihoods of the test sets evaluated at the parameters computed on the training sets, obtaining the so-called \emph{cross-validated} log-likelihood. The constant $c$ is chosen such that the \emph{cross-validated} log-likelihood is maximized. This can be viewed as a function of $c$ only (Smyth, 1996), and would solve the issue of overfitting as training and test sets are independent (Arlot and Celisse, 2010). 

In details, let us partition $K$ times the full data set $\left\lbrace \mathbf{x}_i; i \in N \right\rbrace_n$ into two parts, a training set $\mathbf{x}_{S} = \left\lbrace \mathbf{x}_i ;i \in S \right\rbrace_{n_S},$ and a test set $\mathbf{x}_{\bar{S}} = \left\lbrace \mathbf{x}_i ; i \in \bar{S} \right\rbrace_{n_{\bar{S}}}$ with $S \cup \bar{S} = N$ and $n_S+n_{\bar{S}}=n.$ For the $k$-th partition, let $\hat{\theta}(c,S_{k})$ be the constrained maximum likelihood estimator based on the training set $\mathbf{x}_{S_{k}}.$ Furthermore, let $l_{\bar{S}_{k}} [\hat{\theta}(c,S_k)]$ be the log-likelihood function evaluated at the test set $\mathbf{x}_{\bar{S}_k}.$ The \emph{cross-validated} log-likelihood is defined as the sum of the contributions of each test set to the log-likelihood
\begin{equation}
\text{CV}(c)=\sum_{k=1}^K l_{\bar{S}_k} [\hat{\theta}(c,S_k)].
\end{equation}
The best $c$ is chosen as the maximizer of $\text{CV}(c)$.

Further details on the choice of the number of random partitions $K$ and of the sizes of training and test sets are given in Section \ref{simulation}.

\section{Algorithm} \label{alg}
The objective is to maximize \eqref{eq:like} under the constraints \eqref{eq:gnc}.
Thanks to the equivariance property of the constraints, we can act any linear affine transformation to the data. This is useful since it will suffice to transform the data so to have $\boldsymbol{\Psi}=\mathbf{I}_J$ and the existing algorithm of Ingrassia and Rocci (2007) can be applied on the transformed data.

The transformation is $\mathbf{x}^*=\mathbf{L}^{-\frac{1}{2}}\mathbf{Q}^{'}\mathbf{x},$ where $\boldsymbol{\Psi}=\mathbf{QLQ}'$ is the singular value decomposition of $\boldsymbol{\Psi}.$ This leads to $\boldsymbol{\Psi}^*=\mathbf{L}^{-\frac{1}{2}}\mathbf{Q}' \boldsymbol{\Psi}\mathbf{QL}^{-\frac{1}{2}}=\mathbf{I}_J$.

For sake of completeness, we recall briefly the updates of the algorithm proposed by Ingrassia and Rocci (2007).\\
\\
\textbf{Update} $u_{ig}$, $p_g,$ $\boldsymbol{\mu}_g$ \\
As in the case of a normal mixture, the updates are
\begin{equation}
u_{ig}=\frac{p_g  \phi(\mathbf{x}_i;\boldsymbol{\mu}_g,\boldsymbol{\Sigma}_g)}{\sum_{h=1}^G  \phi(\mathbf{x}_i;\boldsymbol{\mu}_h,\boldsymbol{\Sigma}_h)};
\end{equation}
\begin{equation}
p_g=\frac{1}{n}\sum_{i=1}^n u_{ig};
\end{equation}
\begin{equation}
\mu_g=\frac{\sum_i u_{ig}\mathbf{x}_i}{\sum_i u_{ig}}.
\end{equation}
\textbf{Update} $\boldsymbol{\Sigma}_g$\\ 
Compute 
\begin{equation}
\mathbf{S}_g=\frac{1}{\sum_i u_{ig}}\sum_i u_{ig}(\mathbf{x}_i-\mu_g)(\mathbf{x}_i-\mu_g)',
\end{equation}
and set
\begin{equation}
\lambda_{qg}=\min\left( \frac{1}{\sqrt{c}},\max\left(\sqrt{c},l_{qg} \right) \right), 
\end{equation}
where $\mathbf{L}_g=\textnormal{diag}\left(l_{1g},\dots l_{Jg} \right)$ is the diagonal matrix of the eigenvalues in non decreasing order of $\mathbf{S}_g,$ and $\mathbf{S}_g=\mathbf{Q}_g \mathbf{L}_g \mathbf{Q}_g'$ its singular value decomposition.
Letting $\boldsymbol{\Lambda}_g=\textnormal{diag}\left(\lambda_{1g},\dots \lambda_{jg} \right)$, the update of $\boldsymbol{\Sigma}_g$ is given by
\begin{equation}
\boldsymbol{\Sigma}_g=\mathbf{Q}_g\boldsymbol{\Lambda}_g \mathbf{Q}_g'.
\end{equation}

\section{Simulation study}\label{simulation}
\subsection{Design}
In this section we perform a simulation experiment in order to compare the performance of the proposed methods with respect to some existing approaches in the literature. In particular we consider the following seven algorithms:
\begin{enumerate}
\item Unconstrained
\begin{enumerate}
\item homoscedastic normal (homN), within covariance matrix $\boldsymbol{\Sigma};$ 
\item heteroscedastic normal (hetN), $0.0000001 \leq \lambda_j(\boldsymbol{\Sigma}_g) \leq 10000000$ to prevent degeneracy and numerical instability;
\item homoscedastic Student \textit{t} (hom\textit{t}), scale matrix $\boldsymbol{\Xi},$ $\beta=4$ (McLachlan and Peel, 1998).
\end{enumerate}
\item Constrained
\begin{enumerate}
\item sample covariance (con$\mathbf{S}$), $\boldsymbol{\Psi}=\mathbf{S};$
\item normal (conN), $\boldsymbol{\Psi}=\boldsymbol{\Sigma};$
\item Student \textit{t} (con\textit{t}), $\boldsymbol{\Psi}=\frac{\beta \boldsymbol{\Xi}}{(\beta -2)}.$
\end{enumerate}
\end{enumerate}
For each sample, we randomly split the data $K = 25$ times into a training set $\mathbf{x}_{S},$ and a test set $\mathbf{x}_{\bar{S}}.$ Choosing how many times to partition the full data set is a trade-off between variability of the estimates and computational burden. As Smyth (2000) argues in the context of model selection for probabilistic clustering using cross-validation, the larger the value of $K,$ the less the variability in the log-likelihood estimates. In practice - the Author argues - values of $K$ between 20 and 50 appear adequate for most applications.

The choice of the size of the test set must be such that the training set has all components represented. If one component is not represented in the test set, but the parameters are correctly estimated using the training set, the test set log-likelihood will correctly display the fit of the model. By contrast, if one component is not represented in the training set, although estimation of the other components’ parameters can be correct, the fit displayed by the test set log-likelihood will be poor. Van der Laan, Dudoit, and Keles (2004) found, in their simulation study, that the likelihood-based cross-validation procedure is performing equally well with any choice of the relative size of the test set between 0.1 and 0.5. As argued in Kearns (1996), the importance of choosing an optimal size for the training set increases as the target function becomes more complex relative to the sample size. Bearing this in mind, we choose to consider a training set of size $n_S=n - \frac{n}{10},$ and a test set $\mathbf{x}_{\bar{S}}$ of size $n_{\bar{S}}=\frac{n}{10}.$

Then the cross-validation scheme, as described in Section \ref{crossvalid}, is applied and the optimal $c$ is chosen by using a line search with six function evaluations. 
The sample data have been generated from $G$-class mixtures of heteroscedastic $J$-variate normal distributions with:
\begin{itemize}
\item $n=50,$ $100,$ $200;$
\item $J=5,$ $8;$
\item prior membership probabilities $\mathbf{p}=(0.2, 0.3, 0.5)',$ $(0.1, 0.4, 0.5)',$ $(0.1, 0.1, 0.2, 0.3, 0.3)'.$
\end{itemize}
This yields a total of $2 \times 3 \times 3$ simulation conditions.

For each simulation condition, we generate 250 data sets, each with different means and covariance matrices, where:
\begin{itemize}
\item component means $\mu_{jg} \sim N(0,1.5^2)$, independent;
\item eigenvalues of the covariance matrices $\lambda_{jg} \sim U(0,\frac{g}{\text{sep}})$, independent with $\text{sep}=2$;
\item eigenvectors of the covariance matrices generated by orthonormalizing matrices generated independently from a standard normal distribution.
\end{itemize}

It is well known that the EM for GMM is sensitive to the initial position, especially in the multivariate context (among others, McLachlan and Peel, 2000). We choose to adopt the standard \emph{multiple random starts} strategy. That is, for each data set, 10 random initial partitions are generated: these are used as starting values for the M-step of all the seven algorithms under analysis. For conN, conS, and cont, a constrained algorithm with arbitrary lower and upper bounds of respectively  $0.5$ and $2$ is run in order to exclude degenerate (and some spurious) solutions, and the estimated clustering is used to initialize the cross-validation scheme. The alternative option of directly generating 10 different starts for each training set - within the cross-validation scheme - would have added little in terms of accuracy of the final estimates.

Concerning the root selection criterion, for the unconstrained algorithm, we select the roots yielding the highest likelihood, whereas for the constrained algorithms we select the roots based on the cross-validated likelihood.   

The performance of the different techniques has been analyzed in terms of:
\begin{itemize}
\item MAD (Mean absolute deviation): $\sum_{g=1}^G \sum_{i=1}^n |p(g|x_i)-\hat{p}(g|x_i)|;$
\item ARand (Adjusted Rand index; Hubert and Arabie, 1985);
\item computational time needed to analyze a single data set;
\item the value of the calibrated constant $c$ (for the constrained approach only).
\end{itemize}
The MAD is computed evaluating the above expression for all possible permutations of the estimated classes. The final MAD reported refers to the permutation which yields the lowest difference, and measures inaccuracy of estimated \emph{fuzzy} classification - whereas ARand measures accuracy of estimated \emph{crisp} classification.

In addition, we tested the robustness of the results with respect to changes in 1) the cross-validation settings, and 2) the level of class separation. In order to test robustness with respect to cross-validation settings, we considered a subset of the above simulation conditions as follows. 250 samples, of 50, 100, and 200 observations, were generated from a 3-group 8-variate heteroscedastic Gaussian mixture model, with prior class membership probabilities of 0.1, 0.4, and 0.5.

The same setting was used in order to count how many different local maxima each algorithm converged to over the 10 random initializations considered. This serves the purpose of providing some information on the likelihood surface.

Class separation has been manipulated by controlling the dispersion of the group conditional covariance matrices' eigenvalues (through the above sep value): higher dispersion levels correspond to overlap between the classes. Considering the above full simulation as corresponding to fixed moderate separation ($\text{sep}=2$), this final setup compares results for low, moderate, and high separation levels - respectively $\text{sep}=1,$ $\text{sep}=2,$ and $\text{sep}=3$. The subset of simulation conditions considered is as follows. 250 samples, of 50 observations each, were generated from a 3-group and 5-group heteroscedastic Gaussian mixture model, with prior class membership probabilities of respectively 0.2, 0.3, and 0.5; 0.1, 0.4, and 0.5; 0.1, 0.1, 0.2, 0.3, and 0.3. Table \ref{tabsimcon} summarizes the conditions explored in all testing setups.

\FloatBarrier
\begin{table}[h!]
	\centering
		\begin{tabular}{lcccccccc}
		\hline \hline
	&	Full simulation	&	Cross-val settings	&	Class-sep	&	N. local max	\\
	&		&		&		&		\\
$J=5$, $p=(0.2,0.3,0.5)'$, $n=50$	&	\checkmark	&	$\times$	&	\checkmark	&	$\times$	\\
$J=5$, $p=(0.2,0.3,0.5)'$, $n=100$	&	\checkmark	&	$\times$	&	$\times$	&	$\times$	\\
$J=5$, $p=(0.2,0.3,0.5)'$, $n=200$	&	\checkmark	&	$\times$	&	$\times$	&	$\times$	\\
$J=5$, $p=(0.1,0.4,0.5)'$, $n=50$	&	\checkmark	&	$\times$	&	\checkmark	&	$\times$	\\
$J=5$, $p=(0.1,0.4,0.5)'$, $n=100$	&	\checkmark	&	$\times$	&	$\times$	&	$\times$	\\
$J=5$, $p=(0.1,0.4,0.5)'$, $n=200$	&	\checkmark	&	$\times$	&	$\times$	&	$\times$	\\
$J=5$, $p=(0.1,0.1,0.2,0.3,0.3)'$, $n=50$	&	\checkmark	&	$\times$	&	\checkmark	&	$\times$	\\
$J=5$, $p=(0.1,0.1,0.2,0.3,0.3)'$, $n=100$	&	\checkmark	&	$\times$	&	$\times$	&	$\times$	\\
$J=5$, $p=(0.1,0.1,0.2,0.3,0.3)'$, $n=200$	&	\checkmark	&	$\times$	&	$\times$	&	$\times$	\\
$J=8$, $p=(0.2,0.3,0.5)'$, $n=50$	&	\checkmark	&	$\times$	&	\checkmark	&	$\times$	\\
$J=8$, $p=(0.2,0.3,0.5)'$, $n=100$	&	\checkmark	&	$\times$	&	$\times$	&	$\times$	\\
$J=8$, $p=(0.2,0.3,0.5)'$, $n=200$	&	\checkmark	&	$\times$	&	$\times$	&	$\times$	\\
$J=8$, $p=(0.1,0.4,0.5)'$, $n=50$	&	\checkmark	&	\checkmark	&	\checkmark	&	\checkmark	\\
$J=8$, $p=(0.1,0.4,0.5)'$, $n=100$	&	\checkmark	&	\checkmark	&	$\times$	&	\checkmark	\\
$J=8$, $p=(0.1,0.4,0.5)'$, $n=200$	&	\checkmark	&	\checkmark	&	$\times$	&	\checkmark	\\
$J=8$, $p=(0.1,0.1,0.2,0.3,0.3)'$, $n=50$	&	\checkmark	&	$\times$	&	\checkmark	&	$\times$	\\
$J=8$, $p=(0.1,0.1,0.2,0.3,0.3)'$, $n=100$	&	\checkmark	&	$\times$	&	$\times$	&	$\times$	\\
$J=8$, $p=(0.1,0.1,0.2,0.3,0.3)'$, $n=200$	&	\checkmark	&	$\times$	&	$\times$	&	$\times$	\\

		\hline \hline
		\end{tabular}
	\caption{Cross-table of simulation condition and simulation type.}
	\label{tabsimcon}
\end{table}

 \FloatBarrier

\subsection{Simulation results}
Tables \ref{tabfullj5} and \ref{tabfullj8} present the results obtained with, respectively, $J=5$ and $J=8.$

Among the unconstrained approaches, as expected, for small samples ($n = 50$), the heteroscedastic
normal (hetN) performs poorly, while the homoscedastic Student-t (homt) works nicely. However, the constrained approach (cont) is able to cope with such a small sample size and to improve the performance of the unconstrained approach. The heteroscedastic performs poorly, especially with higher model complexity in terms of number of components and variables relative to the sample size. A similar pattern is, indeed, observed for $n=100$ when $J=5$ with $G=5$ and when $J=8$. 
As the sample size gets larger ($n = 200$), the homoscedastic models are the worst, while the unconstrained heteroscedastic increases in quality of classification. Interestingly, even for a large sample size, conS, conN and cont yield higher or equal quality estimation compared to the unconstrained approach. However, on average, cont seems to be the best, especially for small sample sizes. When $G=3$, the gains observed by the constrained approaches, in terms of cluster recovery, are more pronounced in presence of a component with a small weight ($p=(0.1,0.4,0.5)'$). In general we observe that, whereas increasing the sample size improves the performance of all methods, an increasing number of components lowers the quality of the clustering results. 

\begin{figure}
\centering
\includegraphics[clip, trim=0cm 8.75cm 0cm 2.5cm,width=\linewidth]{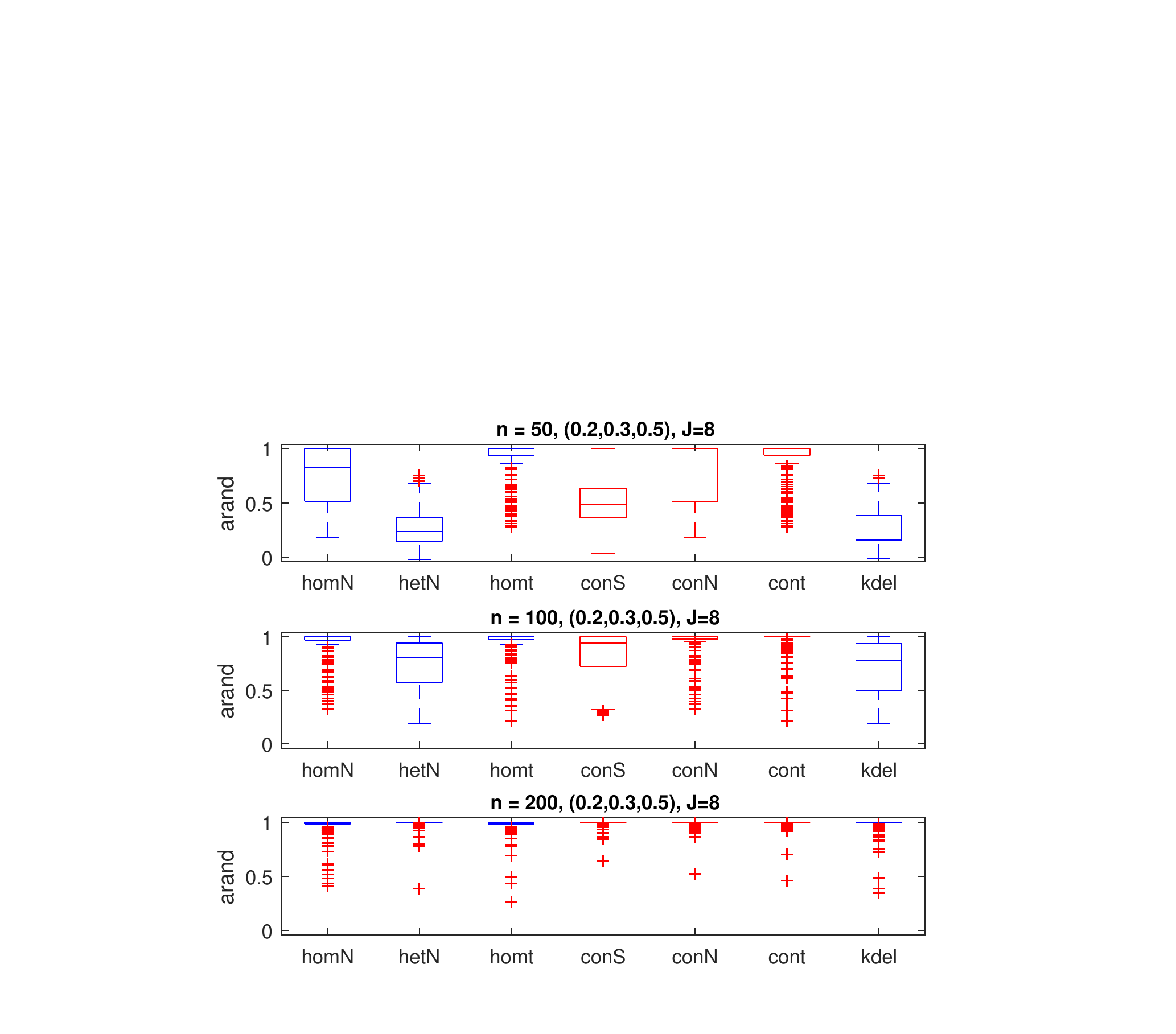}
\caption{Boxplot of the ARand values observed in $250$ simulated data sets, $n=50,$ $n=100,$ and $n=200,$ $p=(0.2,0.3,0.5)',$ and $J=8.$}
\label{fig235}
\end{figure}

\begin{figure}
\centering
\includegraphics[clip, trim=0cm 8.75cm 0cm 2.5cm,width=\linewidth]{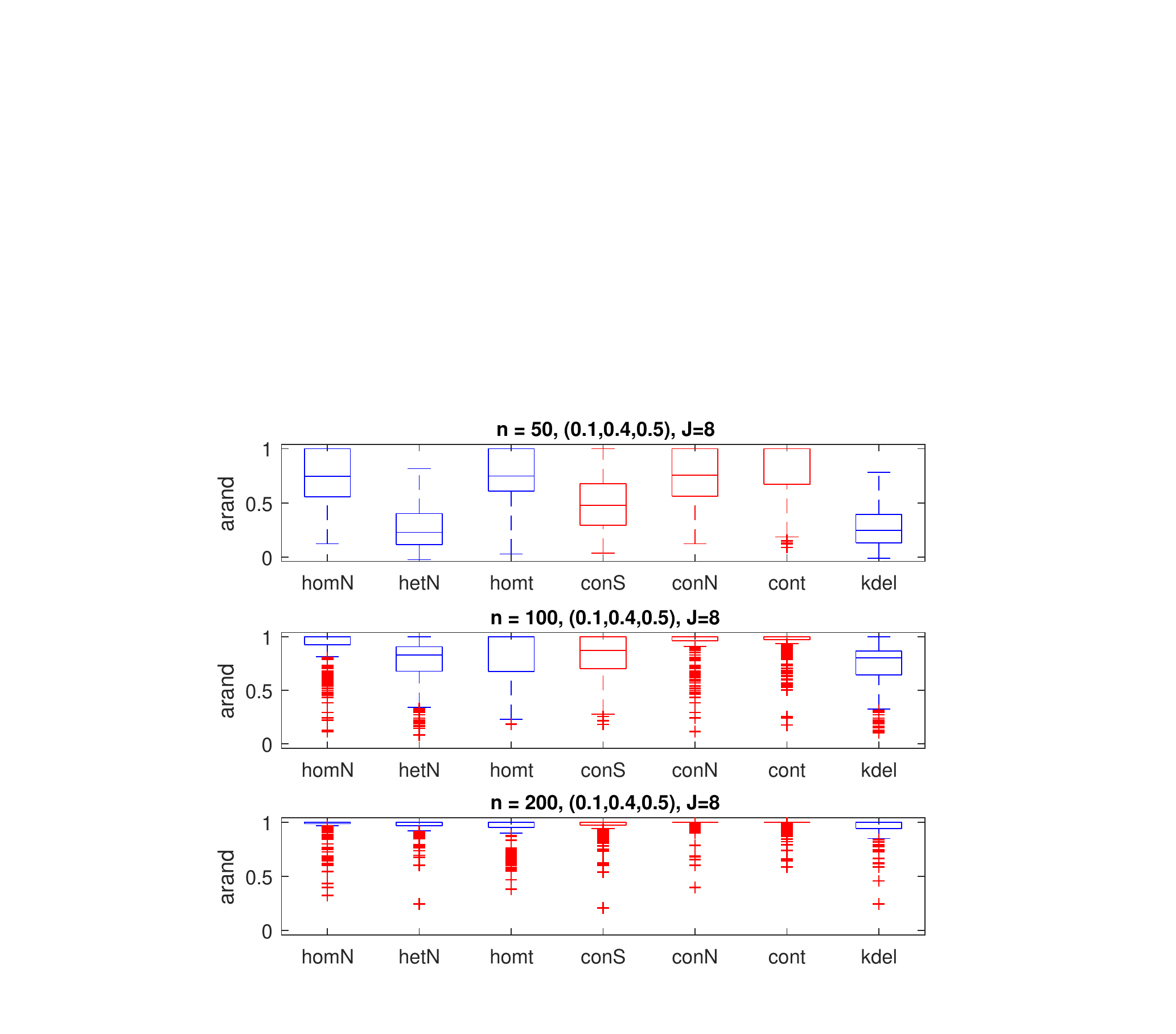}
\caption{Boxplot of the ARand values observed in $250$ simulated data sets, $n=50,$ $n=100,$ and $n=200,$ $p=(0.1,0.4,0.5)',$ and $J=8.$}
\label{fig145}
\end{figure}

\begin{figure}
\centering
\includegraphics[clip, trim=0cm 8.75cm 0cm 2.5cm,width=\linewidth]{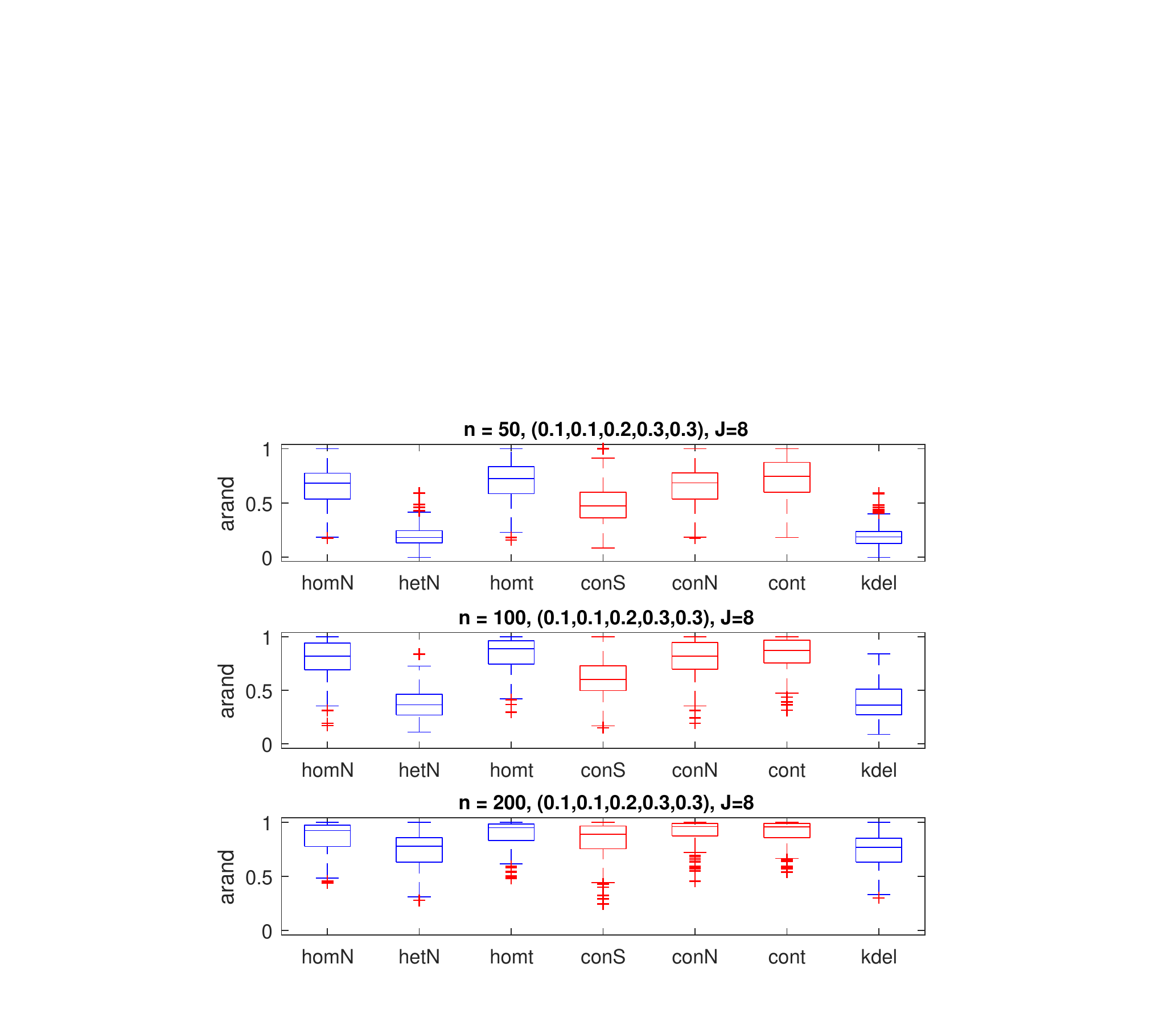}
\caption{Boxplot of the ARand values observed in $250$ simulated data sets, $n=50,$ $n=100,$ and $n=200,$ $p=(0.1,0.1,0.2,0.3,0.3)',$ and $J=8.$}
\label{fig11233}
\end{figure}

The results point out that hetN and conS suffer higher values of $J.$ This is not surprising, as a growing number of variables causes, all else equal, parameter proliferation and a consequent loss in the accuracy of the estimation of the class conditional covariance matrices. Although parameter proliferation is limited with conS, this is constructed based on the choice of the sample covariance matrix as target, which is also very sensible to an increasing $J$ (holding $n$ fixed). In addition, the sample covariance matrix is the sum of the within and the between variance: as such it does not seem to be the best choice for $\boldsymbol{\Psi}.$ On the other hand, homN, homt, conN and cont process the increase in $J$ from 5 to 8 as additional information that, all else equal, improves the quality of the estimation (see Table \ref{tabfullj8}, and Figures \ref{fig235}, \ref{fig145}, and \ref{fig11233}. This is typically the case in finite mixtures with discrete variables (among others, Vermunt, 2010; Di Mari, Oberski, and Vermunt, 2016).

Overall the results show that the constant $c$ decreases in all constrained approaches as the sample size increases, coherently with the results of Xu et al. (2010) in the univariate case. 

In terms of computational time, even if the sample covariance matrix is faster to compute, simulation results show that conS converges slower than conN and cont.
\FloatBarrier
\begin{table}[h!]
	\centering
		\begin{tabular}{cclcccccc}
		\hline \hline
	&		&		&	homN	&	hetN	&	homt	&	conS	&	conN	&	cont	\\ \hline 
	&		&		&		 	&			&		   &		&		&				\\
p=(0.2,0.3,0.5)'	&	n=50	&	MAD	&	0.11	&	0.25	&	0.11	&	0.16	&	0.11	&	0.09	\\
	&		&	ARand	&	0.78	&	0.52	&	0.79	&	0.68	&	0.79	&	0.82	\\
	&		&	time	&	0.10	&	0.08	&	0.09	&	1.17	&	0.67	&	0.73	\\
	&		&	c	&		&		&		&	0.32	&	0.93	&	0.79			\\
	&		&		&		&		&		&		&		&		\\
	&	n=100	&	MAD	&	0.08	&	0.07	&	0.06	&	0.06	&	0.06	&	0.05	\\
	&		&	ARand	&	0.84	&	0.86	&	0.87	&	0.87	&	0.88	&	0.91	\\
	&		&	time	&	0.19	&	0.14	&	0.14	&	1.22	&	0.81	&	0.85	\\
	&		&	c	&		&		&		&	0.12	&	0.77	&	0.53			\\
	&		&		&		&		&		&		&		&		\\
	&	n=200	&	MAD	&	0.06	&	0.02	&	0.05	&	0.02	&	0.02	&	0.02	\\
	&		&	ARand	&	0.88	&	0.96	&	0.90	&	0.96	&	0.96	&	0.96	\\
	&		&	time	&	0.42	&	0.21	&	0.21	&	1.59	&	1.11	&	1.16	\\
	&		&	c	&		&		&		&	0.07	&	0.39	&	0.28	\\ \hline
	&		&		&		&		&		&		&		&		\\
p=(0.1,0.4,0.5)'	&	n=50	&	MAD	&	0.12	&	0.23	&	0.17	&	0.19	&	0.12	&	0.12	\\
	&		&	ARand	&	0.77	&	0.54	&	0.72	&	0.65	&	0.78	&	0.79	\\
	&		&	time	&	0.11	&	0.08	&	0.10	&	1.2	&	0.70	&	0.77	\\
	&		&	c	&		&		&		&	0.34	&	0.93	&	0.78			\\ 
	&		&		&		&		&		&		&		&		\\
	&	n=100	&	MAD	&	0.09	&	0.09	&	0.15	&	0.09	&	0.07	&	0.06	\\
	&		&	ARand	&	0.82	&	0.83	&	0.76	&	0.83	&	0.86	&	0.89	\\
	&		&	time	&	0.19	&	0.15	&	0.17	&	1.28	&	0.84	&	0.91	\\
	&		&	c	&		&		&		&	0.14	&	0.79	&	0.52			\\ 
	&		&		&		&		&		&		&		&		\\
	&	n=200	&	MAD	&	0.07	&	0.03	&	0.11	&	0.03	&	0.03	&	0.03	\\
	&		&	ARand	&	0.87	&	0.95	&	0.82	&	0.93	&	0.94	&	0.94	\\
	&		&	time	&	0.47	&	0.25	&	0.35	&	1.78	&	1.28	&	1.38	\\
	&		&	c	&		&		&		&	0.07	&	0.45	&	0.31	\\ \hline
	&		&		&		&		&		&		&		&		\\
p=(0.1,0.1,0.2,0.3,0.3)'	&	n=50	&	MAD	&	0.27	&	0.46	&	0.28	&	0.31	&	0.27	&	0.26	\\
	&		&	ARand	&	0.58	&	0.29	&	0.57	&	0.52	&	0.58	&	0.58	\\
	&		&	time	&	0.12	&	0.09	&	0.15	&	1.49	&	0.99	&	1.05	\\
	&		&	c	&		&		&		&	0.42	&	0.95	&	0.92	\\ 
	&		&		&		&		&		&		&		&		\\
	&	n=100	&	MAD	&	0.23	&	0.35	&	0.21	&	0.26	&	0.21	&	0.19	\\
	&		&	ARand	&	0.65	&	0.48	&	0.68	&	0.60	&	0.67	&	0.70	\\
	&		&	time	&	0.26	&	0.24	&	0.28	&	2.03	&	1.49	&	1.64	\\
	&		&	c	&		&		&		&	0.27	&	0.90	&	0.66	\\ 
	&		&		&		&		&		&		&		&		\\
	&	n=200	&	MAD	&	0.21	&	0.17	&	0.18	&	0.16	&	0.13	&	0.14	\\
	&		&	ARand	&	0.68	&	0.73	&	0.72	&	0.74	&	0.79	&	0.78	\\
	&		&	time	&	0.71	&	0.62	&	0.65	&	3.87	&	2.75	&	2.8	\\
	&		&	c	&		&		&		&	0.13	&	0.53	&	0.40	\\ \hline \hline
\end{tabular}
	\caption{Average results over 250 simulated data sets of five-variate ($J=5$) GMM estimation, with three and five components. Initialization from ten random starts.}
	\label{tabfullj5}
\end{table}
\FloatBarrier

\begin{table}[h!]
	\centering
		\begin{tabular}{cclccccccc}
		\hline \hline
	&		&		&	homN	&	hetN	&	homt	&	conS	&	conN	&	cont	\\ \hline
	&		&		&		&		&		&		&		&		\\
p=(0.2,0.3,0.5)'	&	n=50	&	MAD	&	0.13	&	0.39	&	0.05	&	0.25	&	0.13	&	0.04	\\
	&		&	ARand	&	0.74	&	0.27	&	0.91	&	0.50	&	0.75	&	0.91	\\
	&		&	time	&	0.08	&	0.06	&	0.09	&	1.51	&	0.66	&	0.70	\\
	&		&	c	&		&		&		&	0.58	&	0.94	&	0.84	\\
	&		&		&		&		&		&		&		&		\\
	&	n=100	&	MAD	&	0.03	&	0.12	&	0.02	&	0.07	&	0.02	&	0.01	\\
	&		&	ARand	&	0.95	&	0.75	&	0.96	&	0.84	&	0.96	&	0.97	\\
	&		&	time	&	0.17	&	0.14	&	0.13	&	1.51	&	0.81	&	0.85	\\
	&		&	c	&		&		&		&	0.20	&	0.94	&	0.60	\\
	&		&		&		&		&		&		&		&		\\
	&	n=200	&	MAD	&	0.01	&	0.00	&	0.01	&	0.00	&	0.00	&	0.00	\\
	&		&	ARand	&	0.97	&	0.99	&	0.98	&	0.99	&	0.99	&	0.99	\\
	&		&	time	&	0.41	&	0.21	&	0.24	&	1.71	&	1.14	&	1.18	\\
	&		&	c	&		&		&		&	0.07	&	0.50	&	0.34	\\ \hline
	&		&		&		&		&		&		&		&		\\
p=(0.1,0.4,0.5)'	&	n=50	&	MAD	&	0.15	&	0.37	&	0.14	&	0.27	&	0.14	&	0.10	\\
	&		&	ARand	&	0.72	&	0.28	&	0.77	&	0.49	&	0.73	&	0.83	\\
	&		&	time	&	0.09	&	0.06	&	0.11	&	1.66	&	0.72	&	0.76	\\
	&		&	c	&		&		&		&	0.61	&	0.95	&	0.81	\\
	&		&		&		&		&		&		&		&		\\
	&	n=100	&	MAD	&	0.05	&	0.12	&	0.09	&	0.10	&	0.04	&	0.03	\\
	&		&	ARand	&	0.90	&	0.77	&	0.86	&	0.81	&	0.93	&	0.94	\\
	&		&	time	&	0.18	&	0.15	&	0.17	&	1.62	&	0.88	&	0.95	\\
	&		&	c	&		&		&		&	0.24	&	0.89	&	0.58	\\ 
	&		&		&		&		&		&		&		&		\\
	&	n=200	&	MAD	&	0.02	&	0.02	&	0.05	&	0.02	&	0.01	&	0.01	\\
	&		&	ARand	&	0.96	&	0.97	&	0.92	&	0.97	&	0.98	&	0.98	\\
	&		&	time	&	0.44	&	0.26	&	0.35	&	1.77	&	1.27	&	1.34	\\
	&		&	c	&		&		&		&	0.08	&	0.56	&	0.36	\\ \hline
	&		&		&		&		&		&		&		&		\\
p=(0.1,0.1,0.2,0.3,0.3)'	&	n=50	&	MAD	&	0.22	&	0.52	&	0.19	&	0.33	&	0.21	&	0.18	\\
	&		&	ARand	&	0.65	&	0.19	&	0.70	&	0.49	&	0.66	&	0.72	\\
	&		&	time	&	0.09	&	0.05	&	0.16	&	1.98	&	1.05	&	1.10	\\
	&		&	c	&		&		&		&	0.60	&	0.95	&	0.93	\\ 
	&		&		&		&		&		&		&		&		\\
	&	n=100	&	MAD	&	0.13	&	0.40	&	0.10	&	0.25	&	0.13	&	0.10	\\
	&		&	ARand	&	0.80	&	0.38	&	0.84	&	0.61	&	0.80	&	0.84	\\
	&		&	time	&	0.22	&	0.18	&	0.26	&	2.11	&	1.49	&	1.63	\\
	&		&	c	&		&		&		&	0.37	&	0.93	&	0.69	\\ 
	&		&		&		&		&		&		&		&		\\
	&	n=200	&	MAD	&	0.09	&	0.18	&	0.07	&	0.10	&	0.05	&	0.06	\\
	&		&	ARand	&	0.86	&	0.74	&	0.90	&	0.84	&	0.91	&	0.91	\\
	&		&	time	&	0.57	&	0.58	&	0.52	&	3.46	&	2.44	&	2.40	\\
	&		&	c	&		&		&		&	0.18	&	0.68	&	0.46	\\ \hline \hline
\end{tabular}
	\caption{Average results over 250 simulated data sets of eight-variate ($J=8$) GMM estimation, with three and five components. Initialization from ten random starts.}
	\label{tabfullj8}
\end{table}
\FloatBarrier
 \FloatBarrier

Table \ref{tablocmax} displays the means and the medians of the number of local maxima the methods under comparison found in the reduced simulation setup (see Table \ref{tabsimcon}). We observe that conN and cont, on average, yield the lowest number of local maxima in all the three sample size conditions. Increasing the sample size yields a more stable behavior also for the other methods.

\FloatBarrier
\begin{table}[h!]
	\centering
		\begin{tabular}{clllllll}
		\hline \hline
	&		&	homN	&	hetN	&	homt	&	conS	&	conN	&	cont		\\ \hline
n=50	& 	mean	&	9.64  	&	9.99	&	7.01	&	9.86	&	5.31	&	4.99	\\
	& 	median	&	10  	&	10	&	7	&	10	&	5	&	5	\\
	&		&		&		&		&		&		&		 \\
n=100	& 	mean	&	7.08  	&	9.94	&	4.71	&	8.50	&	3.96	&	4.64	\\
	& 	median	&	7  	&	10	&	5	&	9	&	4	&	4	\\
	&		&		&		&		&		&		&		\\
n=200	& 	mean	&	3.82  	&	7.04	&	3.16	&	3.78	&	3.37	&	3.59	\\
	& 	median	&	4  	&	7	&	3	&	3	&	3	&	3	\\ \hline
\end{tabular}
	\caption{Mean and median number of local maxima, over $250$ simulated data sets of eight-variate ($J=8$) GMM estimation, $G=3$ and $p=(0.1,0.4,0.5)'.$ Initialization from ten random starts.}
	\label{tablocmax}
\end{table}

\FloatBarrier
In Table \ref{tabcrossval} we display results for different cross-validation settings. Perhaps surprisingly, this points out that the results are not sensitive to the choice of the cross-validations settings in terms of classification. Concerning computational time, this is maximal for $n_{\bar{S}}=n/2$ and $K=n,$ whereas it is minimal for $n_{\bar{S}}=n/10$ and $K=n/10.$ Interestingly however, we observe a systematic decrease in $c$ for lower levels of $n_{\bar{S}}:$ this could be due to the fact that accuracy in parameter estimation, with a larger training set, is higher. Hence, the prior information incorporated in the target matrix becomes less important - and the optimal $c$ relatively smaller. 

\FloatBarrier
\begin{table}[h!]
	\centering
		\begin{tabular}{clccccccccccc}
		\hline \hline
			&		&	\multicolumn{3}{c}{$K=n/10$}		&&       \multicolumn{3}{c}{$K=n/5$}		&&	\multicolumn{3}{c}{$K=n$}		\\
\cmidrule{3-5} \cmidrule{7-9} \cmidrule{11-13}																					
	$n_{\bar{S}}=n/2$	&		&	conS	&	conN	&	cont	&&	conS	&	conN	&	cont	&&	conS	&	conN	&	cont	\\ 
			&		&		&		&		&&		&		&		&&		&		&		\\
			&	MAD	&	0.10	&	0.04	&	0.03	&&	0.10	&	0.04	&	0.03	&&	0.10	&	0.04	&	0.03	\\
			&	ARand	&	0.80	&	0.93	&	0.94	&&	0.81	&	0.93	&	0.94	&&	0.81	&	0.93	&	0.94	\\
			&	time	&	0.94	&	0.45	&	0.52	&&	1.29	&	0.68	&	0.73	&&	4.04	&	2.39	&	2.49	\\
			&	c	&	0.25	&	0.89	&	0.58	&&	0.25	&	0.89	&	0.58	&&	0.25	&	0.90	&	0.58	\\
		 	&		&		&		&		&&		&		&		&&		&		&		\\ 																						
		 	&		&		&		&		&&		&		&		&&		&		&		\\ \hline																						
			&		&	\multicolumn{3}{c}{$K=n/10$}		&&       \multicolumn{3}{c}{$K=n/5$}		&&	\multicolumn{3}{c}{$K=n$}		\\
\cmidrule{3-5} \cmidrule{7-9} \cmidrule{11-13}
	$n_{\bar{S}}=n/5$	&		&	conS	&	conN	&	cont	&&	conS	&	conN	&	cont	&&	conS	&	conN	&	cont	\\ 
			&		&		&		&		&&		&		&		&&		&		&		\\
			&	MAD	&	0.10	&	0.03	&	0.04	&&	0.09	&	0.03	&	0.04	&&	0.09	&	0.03	&	0.03	\\
			&	ARand	&	0.81	&	0.94	&	0.93	&&	0.82	&	0.94	&	0.93	&&	0.83	&	0.95	&	0.94	\\
			&	time	&	0.87	&	0.44	&	0.48	&&	1.11	&	0.65	&	0.69	&&	3.15	&	2.33	&	2.39	\\
			&	c	&	0.10	&	0.69	&	0.47	&&	0.10	&	0.70	&	0.47	&&	0.10	&	0.70	&	0.47	\\
		 	&		&		&		&		&&		&		&		&&		&		&		\\ 																						
		 	&		&		&		&		&&		&		&		&&		&		&		\\ \hline																			
			&		&	\multicolumn{3}{c}{$K=n/10$}		&&       \multicolumn{3}{c}{$K=n/5$}		&&	\multicolumn{3}{c}{$K=n$}		\\
\cmidrule{3-5} \cmidrule{7-9} \cmidrule{11-13}
	$n_{\bar{S}}=n/10$	&		&	conS	&	conN	&	cont	&&	conS	&	conN	&	cont	&&	conS	&	conN	&	cont	\\ 
			&		&		&		&		&&		&		&		&&		&		&		\\
			&	MAD	&	0.10	&	0.03	&	0.04	&&	0.09	&	0.03	&	0.04	&&	0.09	&	0.03	&	0.03	\\
			&	ARand	&	0.80	&	0.94	&	0.93	&&	0.82	&	0.95	&	0.94	&&	0.82	&	0.95	&	0.94	\\
			&	time	&	0.82	&	0.43	&	0.48	&&	1.01	&	0.63	&	0.67	&&	2.84	&	2.32	&	2.34	\\
			&	c	&	0.09	&	0.63	&	0.43	&&	0.09	&	0.63	&	0.43	&&	0.09	&	0.63	&	0.43	\\ \hline \hline
\end{tabular}
	\caption{Average results over 250 simulated data sets, each of sample size 100, of eight-variate ($J=8$) GMM constrained estimation, $G=3$ and $p=(0.1,0.4,0.5)',$ for different cross-validation settings. Initialization from ten random starts.}
	\label{tabcrossval}
\end{table}

\FloatBarrier

Table \ref{tabsepar} gives results on three different levels of class separation, for three and five components with small sample size ($n=50$). Both the unconstrained and the constrained methods have a relatively stronger effect on performance when class separation increases from $\text{sep}=1$ to $\text{sep}=2$ than when it increases from $\text{sep}=2$ to $\text{sep}=3$. Among all the approaches considered, whatever the class separation, cont yields on average the most accurate clustering.

\FloatBarrier
\begin{table}[h!]
	\centering
		\resizebox{0.75\hsize}{!}{\begin{tabular}{lllcccccc}
		\hline \hline
$\text{sep}=1$	&		&		&	homN	&	hetN	&	homt	&	conS	&	conN	&	cont	\\ \cmidrule{3-9}
	&	$p=(0.2,0.3,0.5)'$	&		&		&		&		&		&		&		\\	
	&		&	MAD	&	0.22	&	0.40	&	0.13	&	0.32	&	0.22	&	0.13	\\
	&		&	Arand	&	0.57	&	0.23	&	0.73	&	0.42	&	0.57	&	0.74	\\
	&		&	time	&	0.09	&	0.06	&	0.11	&	1.75	&	0.72	&	0.79	\\
	&		&	$c$	&		&		&		&	0.66	&	0.95	&	0.84	\\
	&		&		&		&		&		&		&		&		\\
	&	$p=(0.1,0.4,0.5)'$	&		&		&		&		&		&		&		\\
	&		&	MAD	&	0.23	&	0.41	&	0.22	&	0.33	&	0.23	&	0.20	\\
	&		&	Arand	&	0.55	&	0.22	&	0.62	&	0.41	&	0.55	&	0.64	\\
	&		&	time	&	0.09	&	0.06	&	0.12	&	1.83	&	0.76	&	0.84	\\
	&		&	$c$	&		&		&		&	0.67	&	0.95	&	0.81	\\
	&		&		&		&		&		&		&		&		\\
	&	$p=(0.1,0.1,0.2,0.3,0.3)'$	&		&		&		&		&		&		&		\\
	&		&	MAD	&	0.38	&	0.55	&	0.34	&	0.43	&	0.37	&	0.34	\\
	&		&	Arand	&	0.40	&	0.15	&	0.45	&	0.34	&	0.41	&	0.46	\\
	&		&	time	&	0.11	&	0.05	&	0.19	&	2.37	&	1.13	&	1.22	\\
	&		&	$c$	&		&		&		&	0.69	&	0.95	&	0.94	\\
	&		&		&		&		&		&		&		&		\\ 
\cmidrule{3-9}
$\text{sep}=2$	&		&		&		&		&		&		&		&		\\
	&	$p=(0.2,0.3,0.5)'$	&		&		&		&		&		&		&				\\
	&		&	MAD	&	0.13	&	0.39	&	0.05	&	0.25	&	0.13	&	0.04	\\
	&		&	Arand	&	0.74	&	0.27	&	0.91	&	0.50	&	0.75	&	0.91	\\
	&		&	time	&	0.08	&	0.06	&	0.09	&	1.51	&	0.66	&	0.70	\\
	&		&	$c$	&		&		&		&	0.58	&	0.94	&	0.84	\\
	&		&		&		&		&		&		&		&		\\
	&	$p=(0.1,0.4,0.5)'$	&		&		&		&		&		&		&				\\
	&		&	MAD	&	0.15	&	0.37	&	0.14	&	0.27	&	0.14	&	0.10	\\
	&		&	Arand	&	0.72	&	0.28	&	0.77	&	0.49	&	0.73	&	0.83	\\
	&		&	time	&	0.09	&	0.06	&	0.11	&	1.66	&	0.72	&	0.76	\\
	&		&	$c$	&		&		&		&	0.61	&	0.95	&	0.81	\\
	&		&		&		&		&		&		&		&				\\
	&	$p=(0.1,0.1,0.2,0.3,0.3)'$	&		&		&		&		&		&		&		\\
	&		&	MAD	&	0.22	&	0.52	&	0.19	&	0.33	&	0.21	&	0.18	\\
	&		&	Arand	&	0.65	&	0.19	&	0.70	&	0.49	&	0.66	&	0.72	\\
	&		&	time	&	0.09	&	0.05	&	0.16	&	1.98	&	1.05	&	1.10	\\
	&		&	$c$	&		&		&		&	0.60	&	0.95	&	0.93	\\
	&		&		&		&		&		&		&		&		\\
\cmidrule{3-9}
$\text{sep}=3$	&		&		&		&		&		&		&		&		\\
	&	$p=(0.2,0.3,0.5)'$	&		&		&		&		&		&		&			\\
	&		&	MAD	&	0.10	&	0.37	&	0.03	&	0.22	&	0.09	&	0.03	\\
	&		&	Arand	&	0.81	&	0.31	&	0.94	&	0.56	&	0.81	&	0.95	\\
	&		&	time	&	0.08	&	0.06	&	0.09	&	1.45	&	0.66	&	0.67	\\
	&		&	$c$	&		&		&		&	0.55	&	0.94	&	0.83	\\
	&		&		&		&		&		&		&		&		\\
	&	$p=(0.1,0.4,0.5)'$	&		&		&		&		&		&		&			\\
	&		&	MAD	&	0.12	&	0.36	&	0.13	&	0.24	&	0.11	&	0.06	\\
	&		&	Arand	&	0.77	&	0.29	&	0.79	&	0.54	&	0.78	&	0.89	\\
	&		&	time	&	0.09	&	0.06	&	0.11	&	1.61	&	0.72	&	0.74	\\
	&		&	$c$	&		&		&		&	0.58	&	0.94	&	0.80	\\
	&		&		&		&		&		&		&		&		\\
	&	$p=(0.1,0.1,0.2,0.3,0.3)'$	&		&		&		&		&		&		&		\\
	&		&	MAD	&	0.15	&	0.49	&	0.14	&	0.28	&	0.14	&	0.12	\\
	&		&	Arand	&	0.77	&	0.23	&	0.79	&	0.55	&	0.78	&	0.81	\\
	&		&	time	&	0.09	&	0.05	&	0.15	&	1.81	&	1.04	&	1.08	\\
	&		&	$c$	&		&		&		&	0.55	&	0.95	&	0.93	\\ \hline

\end{tabular}}
	\caption{Average results over 250 simulated data sets, each of sample size 50, of eight-variate ($J=8$) GMM estimation, three and five components, for different class separation levels. Initialization from ten random starts.}
	\label{tabsepar}
\end{table}

\FloatBarrier

\section{Empirical application: the wine data set}\label{wineapp}
In the present Section we evaluate the seven algorithms on the basis of a data set available at \url{http://www.ics.uci.edu/~mlearn/MLRepository.html}. These data are the results of a chemical analysis of three types of wine - \emph{Barolo, Grignolino} and \emph{Barbera} - grown in the same region in Italy. The analysis determined the quantities of 13 constituents found in each of the three types of wines: alcohol, malic acid, ash, alcalinity of ash, magnesium, total phenols, flavanoids, nonflavanoid phenols, proanthocyanins, color intensity, hue, OD280/OD315 of diluted wines, and proline.

All of the six algorithms have been initialized from the same 50 random starts, assuming $G=3.$ The selected solutions are the ones with the highest likelihood. The cross-validation scheme is the same used in the simulation study. Results are shown in Table \ref{wine}.\\

\begin{table}[h!]
	\centering
		\begin{tabular}{lcccccc}
		\hline
		\hline
	&	homN	&	hetN	&	homt	&	conS	&	conN	&	cont	\\ \hline
	&		&		&		&		&		&		\\    
ARand	&	0.92	&	0.39	&	0.87	&	0.54	&	0.92	&	0.93	\\
time	&	2.22	&	1.89	&	1.68	&	23.34	&	11.32	&	10.71	\\
$c$	&		&		&		&	0.45	&	0.89	&	0.57	\\ \hline \hline
\end{tabular}
	\caption{Comparison of the 6 algorithms in terms of ARand, computational time and optimal $c.$}
	\label{wine}
\end{table}
\indent The homoscedasticity assumption seems to fit well the data. The constrained approach conN equals homN in terms of ARand, whereas cont yields an ARand of 0.93, compared to 0.87 of homt. Confirming the results obtained in the simulation study, cont seems to be the most accurate approach among the ones considered in this work. 

Interestingly, however, all of the constrained approaches improve upon the unconstrained heteroscedastic approach.

\section{Discussion}\label{concl}
In this paper we have proposed affine equivariant constraints for the class conditional covariance matrices of multivariate GMM in order to circumvent the well-known issue of degenerate and spurious solutions in ML estimation. Our approach generalizes the sufficient condition for Hathaway (1985)'s constraints to hold as formulated by Ingrassia (2004). Previous constrained approaches lacked affine equivariance and suffered the choice of an optimal finite-sample scale balance ($c$). The setup we propose is such that the class specific covariance matrices are shrunk towards a pre-specified matrix $\boldsymbol{\Psi}.$ We have been able to show that this yields a clustering method which is equivariant with respect to linear affine transformations of the data, provided that $\boldsymbol{\Psi}$ is changed accordingly. 

A natural choice for the shrinkage target matrix, whenever \emph{a priori} information on the covariance structure of the components is not available, seems to be the covariance matrix of a homoscedastic mixture of normals. For a given choice of the target matrix, we let the data decide, through the constant $c$,  how close to the target the final clustering will be. The tuning constant $c$ is chosen by cross-validation. We have also shown that, given a matrix $\boldsymbol{\Psi},$ our constrained ML estimate can be computed by applying the algorithm of Ingrassia and Rocci (2007) to the data appropriately linearly transformed. This allows us to interpret our proposal as a way to decide how to standardize the data before applying Ingrassia (2004)'s constraints.

The validity of the proposal has been assessed through a simulation study and an empirical example. All constrained approaches yield more accurate estimates than the unconstrained ones. More specifically, cont has been shown to be the best among the constrained approaches this work has been concerned with. This is not surprising, since a random vector conditionally distributed as a Gaussian mixture, given random inverse Wishart covariance matrices, has a marginal homoscedastic mixture of Student $t$'s distribution. 

Given an affine transformation of the data, the equivariance property of the method is guaranteed if also $\boldsymbol{\Psi}$ is adapted accordingly. This requires that the methods used to estimate $\boldsymbol{\Psi}$ from the data be also equivariant. This is the case for the sample covariance matrix and the homoscedastic model, for which \ref{eq:equiGauss} and \ref{eq:equiGaussmix} apply. For a homoscedastic mixture of Student $t$'s, this can also be shown expressing each marginal as a combination of a multivariate Gaussian and Gamma random variables. Then affine equivariance results by applying \ref{eq:equiGauss} (Roth, 2013). All in all, different choices of $\boldsymbol{\Psi}$ can as well be considered, according to the data specificity: however, in order for the method to preserve equivariance, also the method used to estimate $\boldsymbol{\Psi}$ from the data has to be equivariant. 

The equivariant method developed in Gallegos and Ritter (2009a; 2009b) and extended in Ritter (2014) requires to obtain all local maxima of the trimmed likelihood. Our method has the virtue of being easily implementable with a minimal extra computational effort, as we have shown in the simulation study and in the empirical example.

There are cases where the clustering model might assume a specific structure on the relationship between the variables, like local independence (within-cluster diagonal matrices). Such a model is not affine equivariant because some (non diagonal) affine transformations on the data might destroy the local independence. In cases like these, the affine equivariance property of the constraints is not required. Yet our approach can be applied using a diagonal matrix as target. This would prevent the likelihood from degenerating, still improving upon the unconstrained algorithm thanks to the cross-validation strategy we have proposed. Clearly, when all variables in a data set are measured in a common scale, non equivariant constraints are a competitive choice.

An additional issue, pointed out by both the simulation study and the empirical example, is the computational time cross-validation requires to select an optimal $c.$ Whether different cross-validation schemes can speed up the constrained routines can be a topic for future research.


\newpage
\begin{thebibliography}{20}

\bibitem{} Anderson T.W., Gupta, S.D. (1963). Some Inequalities on Characteristic Roots of Matrices.
{\em Biometrika, 50, 522-524}.

\bibitem{} Arlot, S., Celisse, A. (2010). A survey of cross-validation procedures for model selection. \emph{Statistics Surveys, 4, 40-79.}

\bibitem{} Biernacki, C., Celeux, G., Govaert, G. (2003). Choosing starting values for the EM algorithm for getting the highest likelihood in multivariate Gaussian mixture models. \emph{Computational Statistics and Data Analysis, 41(3), 561-575.}

\bibitem{} Biernacki C., Chr\'etien S. (2003). Degeneracy in the maximum likelihood estimation of univariate Gaussian mixtures with the EM.
{\em Statistics and Probability Letters, 61, 373-382}.

\bibitem{} Browne, R. P., Subedi, S., McNicholas, P. (2013). Constrained Optimization for a Subset of the Gaussian Parsimonious Clustering Models. \emph{arXiv preprint arXiv:1306.5824.}

\bibitem{} Chen, J., Tan, X., Zhang, R. (2008). Inference for normal mixtures in mean and variance. \emph{Statistica Sinica, 18(2), 443.}

\bibitem{chen2} Chen J., Tan X. (2009). Inference for multivariate normal mixtures. \emph{Journal of the Multivariate Analysis, 100, 1367-1383.}

\bibitem{ciuperca} Ciuperca, G., Ridolfi, A., Idier, J. (2003). Penalized maximum likelihood estimator for normal mixtures. \emph{Scandinavian Journal of Statistics, 30(1), 45-59.}

\bibitem{} Dawid, A. P. (1981). Some matrix-variate distribution theory: notational considerations and a Bayesian application. \emph{Biometrika, 68(1), 265-274.}

\bibitem{} Dickey, J. M. (1967). Matricvariate generalizations of the multivariate t distribution and the inverted multivariate t distribution. \emph{The Annals of Mathematical Statistics, 38(2), 511-518.}

\bibitem{}
Day N.E. (1969). Estimating the components of a mixture of two normal distributions, 
{\em Biometrika, 56, 463-474}.

\bibitem{}
Di Mari, R., Oberski, D.L., Vermunt, J.K. (2016). Bias-Adjusted Three-Step Latent Markov Modeling With Covariates. \emph{Structural Equation Modeling: A Multidisciplinary Journal}. DOI:10.1080/10705511.2016.1191015.

\bibitem{}
Doherty, K.A.J., Adams, R.G.(2007). 
Unsupervised learning with normalised data and non-euclidean norms. 
{\em Applied Soft Computing 7: 203–21}.

\bibitem{fraley} Fraley, C., Raftery, A. E. (2007). Bayesian regularization for normal mixture estimation and model-based clustering. \emph{Journal of Classification, 24(2), 155-181}.

\bibitem{fritz} Fritz, H., Garcia-Escudero, L.A., Mayo-Iscar, A. (2013). A fast algorithm for robust constrained clustering. \emph{Computational Statistics and Data Analysis,61, 124-136.}

\bibitem{gallegos1} Gallegos, M. T., Ritter, G. (2009a). Trimming algorithms for clustering contaminated grouped data and their robustness. \emph{Advances in Data Analysis and Classification, 3(2), 135-167.}

\bibitem{gallegos2} Gallegos, M. T., Ritter, G. (2009b). Trimmed ML estimation of contaminated mixtures. \emph{Sankhya: The Indian Journal of Statistics, Series A (2008-), 164-220.}

\bibitem{garciaesc} Garcia-Escudero, L.A., Gordaliza, A., Matran, C., Mayo-Iscar, A. (2008). A general trimming approach to robust cluster analysis. \emph{Annals of Statistics,36, 1324-1345}.

\bibitem{garciaesc1} Garcia-Escudero, L. A., Gordaliza, A., Matran, C., Mayo-Iscar, A. (2014). Avoiding spurious local maximizers in mixture modeling. \emph{Statistics and Computing, 25(3), 619-633.}

\bibitem{greselin} Greselin, F., Ingrassia, S. (2013). Maximum likelihood estimation in constrained parameter spaces for mixtures of factor analyzers. \emph{Statistics and Computing, 25(2), 215-226.}

\bibitem{}
Hathaway R. J. (1985). A constrained formulation of maximum-likelihood estimation for normal mixture distributions,
{\em The Annals of Statistics, 13, 795-800}.

\bibitem{}
Hubert L., Arabie P. (1985). Comparing partitions. {\em Journal of Classification, 2, 193-218}.


\bibitem{} 
Ingrassia S. (2004).  A likelihood-based constrained algorithm for multivariate normal mixture models,
 {\em Statistical Methods and Applications, 13, 151-166}.

\bibitem{ing07} 
Ingrassia S., Rocci, R. (2007). A constrained monotone EM algorithm  for finite mixture of multivariate Gaussians.
{\em Computational Statistics and Data Analysis,  51, 5339-5351}.

\bibitem{ing11} Ingrassia, S., Rocci, R. (2011). Degeneracy of the EM algorithm for the MLE of multivariate Gaussian mixtures and dynamic constraints. \emph{Computational Statistics and Data Analysis, 55(4), 1715-1725.}

\bibitem{} James, W., Stein, C. (1961). Estimation with quadratic loss. In \emph{Proceedings of the fourth Berkeley symposium on mathematical statistics and probability (Vol. 1, No. 1961, pp. 361-379).}

\bibitem{}
Kearns, M. (1997). A bound on the error of cross validation using the approximation and estimation rates, with consequences for the training-test split. \emph{Neural Computation, 9(5), 1143-1161.}

\bibitem{}
Kiefer J., Wolfowitz J. (1956). 
Consistency of the maximum likelihood estimator in the presence of infinitely many incidental parameters. {\em Annals of Mathematical Statistics 27, 886–906}.

\bibitem{}
Kiefer, N. M. (1978)
Discrete parameter variation: Efficient estimation of a switching regression model.
{\em Econometrica 46, 427-434}.

\bibitem{} Kim, D., Seo, B. (2014). Assessment of the number of components in Gaussian mixture models in the presence of multiple local maximizers. \emph{Journal of Multivariate Analysis, 125, 100-120.}

\bibitem{}
Kleinber J. (2002). An impossibility theorem for clustering.
In:{\em Advances in Neural Information Processing Systems, (NIPS), 446-453}. MIT Press, Cambridge.

\bibitem{lichman} Lichman, M. (2013). UCI Machine Learning Repository \url{http://archive.ics.uci.edu/ml}. Irvine, CA: University of California, School of Information and Computer Science.


\bibitem{} McLachlan, G. J., Peel, D. (1998). Robust cluster analysis via mixtures of multivariate t-distributions. In \emph{Advances in pattern recognition, 658-666}. Springer Berlin Heidelberg.

\bibitem{}
 McLachlan G.J., Peel D. (2000). {\em Finite Mixture Models}. John Wiley and Sons, New York.

\bibitem{} 
Milligan G.W., Cooper M.C. (1988). A study of standardization of variables in cluster analysis.
{\em Journal of Classification,  5, 181-204}.


\bibitem{} Peel, D., McLachlan, G. J. (2000). Robust mixture modelling using the t distribution. \emph{Statistics and computing, 10(4), 339-348}.

\bibitem{policello} Policello II, G. E. (1981). Conditional maximum likelihood estimation in Gaussian mixtures. In \emph{Statistical Distributions in Scientific Work}, 111-125. Springer Netherlands.

\bibitem{ridolfi1} Ridolfi A., Idier, J. (1999). Penalized maximum likelihood estimation for univariate normal mixture distributions. In \emph{Actes du 17' colloque GRETSI, 259-262}, Vannes, France. 

\bibitem{ridolfi2} Ridolfi A., Idier J. (2000). Penalized maximum likelihood estimation for univariate normal mixture distributions. \emph{Bayesian inference and maximum entropy methods}, MaxEnt Workshops. Gif-sur-Yvette, France, July 2000. 

\bibitem{} Ritter, G. (2014). \emph{Robust cluster analysis and variable selection.} CRC Press.

\bibitem{} Roth, M. (2013). On the multivariate $t$ distribution. \emph{Technical report,} Link\"{o}ping university, Division of automatic control.

\bibitem{seo} Seo B., Kim D. (2012). Root selection in normal mixture models. \emph{Computational Statistics and Data Analysis, 56, 2454--2470}.

\bibitem{} Smyth, P. (1996). \emph{Clustering using Monte-Carlo cross validation.} In \emph{Proceedings of the Second International Conference on Knowledge Discovery and Data Mining, Menlo Park, CA, AAAI Press, pp. 126–133.}

\bibitem{}
Smyth, P. (2000). Model selection for probabilistic clustering using cross-validated likelihood. \emph{Statistics and Computing, 10(1), 63-72.}

\bibitem{snoussi} Snoussi H., Mohammad-Djafari A. (2001). \emph{Penalized maximum likelihood for multivariate Gaussian mixture.} In R.L. Fry Editor \emph{MaxEnt Workshops: Bayesian Inference and
Maximum Entropy Methods, 36-46, Aug. 2001.}

\bibitem{} 
Tan X., Chen J., Zhang R. (2007).
Consistency of the constrained maximum likelihood estimator in finite normal mixture models.
Proceedings of the American Statistical Association, American Statistical Association, Alexandria, VA, 2007, pp. 2113-2119. [CD-ROM].


\bibitem{}
Tanaka K., Takemura A. (2006)
Strong consistency of the maximum likelihood estimator for finite mixtures of location–scale distributions when the scale parameters are exponentially small.
{\em Bernoulli, 12 (6), 1003-1017}.

\bibitem{}
van der Laan, M. J., Dudoit, S., Keles, S. (2004). Asymptotic optimality of likelihood-based cross-validation. \emph{Statistical Applications in Genetics and Molecular Biology, 3(1), 1-23.}

\bibitem{verm1}Vermunt, J. K. (2010). Latent class modeling with covariates: Two improved three-step approaches. \emph{Political analysis, 18}(4), 450--469.

\bibitem{} 
Xu J., Tan X., Zhang R. (2010)
A note on Phillips (1991): ``A constrained maximum likelihood approach to
estimating switching regressions''.
{\em Journal of Econometrics, 154, 35-41}.




\end{thebibliography}
\end{document}